\documentclass[reqno,12pt]{amsart}

\usepackage{fullpage}
\usepackage{amsfonts}
\usepackage{amssymb}
\usepackage{amsmath}
\usepackage{graphicx}
\usepackage{float}

\begin{document}

\newtheorem{prop}{Proposition}
\newtheorem{thm}{Theorem}
\newtheorem{cor}{Corollary}
\newtheorem{lem}{Lemma}
\newtheorem{rem}{Remark}
\newtheorem{defn}{Definition}

\def\Ref#1{Ref.\cite{#1}}

% abrevs
\def\a{\alpha}
\def\b{\beta}
\def\c{\gamma}
\def\d{\kappa}
\def\e{\sigma}
\def\f{\nu}
\def\w{\omega}
\def\v{\theta}

\def\V{U}
\def\H{A}
\def\Y{\Phi}
\def\z{\xi}
\def\E{E}

\def\Rnum{{\mathbb R}}
\def\sgn{{\rm sgn}}
\def\abs#1{|#1|}
\def\diag{{\rm diag}}
\def\Re{{\rm Re}}
\def\t{\dagger}

\def\sech{{\rm sech}}
\def\arctanh{{\rm arctanh}}
\def\arccosh{{\rm arccosh}}
\def\const{\text{const.}}

\def\X{\mathbf{X}}
\def\pr{{\rm pr}}

\def\Lop{{\mathcal L}}
\def\Vop{{\mathcal V}}
\def\Iop{{\mathcal I}}
\def\Jop{{\mathcal J}}
\def\Rop{{\mathcal R}}

\def\crit{\text{crit.}}
\def\side{\text{side}}
\def\nonlin{\text{nonlin.}}
\def\lin{\text{lin.}}

\numberwithin{equation}{section}

\tolerance=50000
\allowdisplaybreaks[4]

\title{Exact solitary wave solutions\\ for a coupled \lowercase{g}KdV-Schrodinger system\\ by a new ODE reduction method}

\author{
Stephen C. Anco$^1$,  
James Hornick$^2$,\\
Sicheng Zhao$^3$,
Thomas Wolf$^1$
\\\\
${}^1$D\lowercase{\scshape{epartment}} \lowercase{\scshape{of}} M\lowercase{\scshape{athematics and}} S\lowercase{\scshape{tatistics}}\\
B\lowercase{\scshape{rock}} U\lowercase{\scshape{niversity}}\\
S\lowercase{\scshape{t.}} C\lowercase{\scshape{atharines}}, O\lowercase{\scshape{ntario}}, C\lowercase{\scshape{anada}} \\
\\
${}^2$D\lowercase{\scshape{epartment}} \lowercase{\scshape{of}} M\lowercase{\scshape{athematics and}} S\lowercase{\scshape{tatistics}}\\
M\lowercase{\scshape{c}}M\lowercase{\scshape{aster}} U\lowercase{\scshape{niversity}}\\
H\lowercase{\scshape{amilton}}, O\lowercase{\scshape{ntario}}, C\lowercase{\scshape{anada}} \\
\\
${}^3$D\lowercase{\scshape{epartment}} \lowercase{\scshape{of}} M\lowercase{\scshape{athematics and}} S\lowercase{\scshape{tatistics}}\\
Q\lowercase{\scshape{ueen's}} U\lowercase{\scshape{niversity}}\\
K\lowercase{\scshape{ingston}}, O\lowercase{\scshape{ntario}}, C\lowercase{\scshape{anada}} \\
}

\begin{abstract}
A new method is developed for finding exact solitary wave solutions of 
a generalized Korteweg-de Vries equation with $p$-power nonlinearity 
coupled to a linear Schr\"odinger equation
arising in many different physical applications. 
This method yields 22 solution families, with $p=1,2,3,4$. 
No solutions for $p>1$ were known previously in the literature. 
For $p=1$, 
four of the solution families contain bright/dark Davydov solitons of the 1st and 2nd kind,
obtained in recent literature by basic ansatze applied to the ODE system for travelling waves.
All of the new solution families have interesting features, 
including bright/dark peaks 
with (up to) $p$ symmetric pairs of side peaks in the amplitude 
and a kink profile for the nonlinear part in the phase. 
The present method is fully systematic and involves several novel steps 
which reduce the travelling wave ODE system to a single nonlinear base ODE 
for which all polynomial solutions are found by symbolic computation. 
It is applicable more generally to other coupled nonlinear dispersive wave equations 
as well as to nonlinear ODE systems of generalized H\'enon-Heiles form. 
\end{abstract}

\maketitle

\section{Introduction}\label{sec:intro}

A wide class of physical wave propagation phenomena are modeled by 
nonlinear dispersive systems. 
The simplest model that combines dispersion and weak nonlinearity is given by 
the Korteweg-de Vries (KdV) equation
$u_t + \a uu_x +\b u_{xxx}=0$ for uni-directional wave motion $u(t,x)$, 
which describes \cite{Cri,DauPey} 
surface waves in shallow water, 
internal waves in stratified fluids, 
ion-acoustic waves in plasmas, 
atmospheric Rossby waves, 
compression waves in granular chains, 
deformations of long-chain molecules and anharmonic lattices,
and numerous other physical applications. 
Higher-order nonlinearities are of physical interest in several of these applications
\cite{Ono, GriPelPol,Mam,MusSha}.

In more complex physical systems, 
nonlinear dispersive wave motion is coupled to high-frequency waves, 
as modeled by a linear Schr\"odinger equation (LS)
$i\psi_t + \d \psi_{xx} = \e u \psi$
for $\psi(t,x)$ in which $u(t,x)$ acts as a potential well. 
Physical applications include 
electron propagation coupled to nonlinear ion-acoustic waves in a collisionless plasma
\cite{NisHojMimIke,Rao}, 
resonant interaction of capillary-gravity waves and internal water waves 
\cite{KawSugKak}, 
electron propagation along deformable molecular chains \cite{Dav-book,DavZol,ZolSpaSav}, 
and energy transport by electrons in anharmonic lattices \cite{CisPel2017,CisPraVilCar2018}.
See \Ref{YosWakKak,GroMal} for further discussion. 

This motivates the study of a general coupled gKdV-LS system
\begin{align}
u_t + \a u^p u_x + \b u_{xxx} & = \c (|\psi|^2)_x, 
\label{u.eqn.phys}
\\
i\psi_t + \d \psi_{xx} & = \e u \psi, 
\label{psi.eqn.phys}
\end{align}
where $p>0$ is the nonlinearity power and $\a$ is the nonlinearity coefficient, 
$\b$ and $\d$ are the dispersion coefficients, 
while $\c$ and $\e$ are the coupling constants. 
All of these parameters are non-zero, real constants. 
Here, $u(t,x)$ is the (real) gKdV field and $\psi$ is the (complex) LS field,
which respectively describe the wave amplitude or lattice deformation, 
and the electron wave function or envelope of high-frequency waves. 

One main question is understanding the nature of 
coherent propagating structures in this system, in particular, 
solitary wave solutions of different shapes and subsonic/supersonic speeds. 
Such solutions have the form of phase-modulated travelling waves
\begin{equation}\label{inv.soln}
u=U(\z),
\quad
\psi=e^{-i\w t}\Psi(\z),
\quad
\z=x-ct, 
\end{equation}
where $c$ is the wave speed, and $\w$ is the frequency. 
In physical applications, 
a travelling wave is supersonic if $c>0$ and subsonic if $c<0$
\cite{DavZol,CisPel2017,GroMal}. 
Here $U$ is a real function and $\Psi$ is a complex function,
which satisfy a coupled nonlinear system of ordinary differential equations (ODEs) 
\begin{equation}\label{ODE.sys}
\b U'''+ (\a U^p-c)U' + \c(|\Psi|^2)' =0,
\quad
\d \Psi'' -ic\Psi'+(\e U +\w)\Psi =0 . 
\end{equation}
Solitary wave solutions require that $U$ decays exponentially for large $|\z|$
to a constant, $b$, 
and likewise that $|\Psi|$ has a similar decay. 

For a KdV nonlinearity, $p=1$,
some exact solutions are known 
which were obtained by use of elementary ODE methods combined with trial ansatze
\cite{NisHojMimIke,CisPel2017}
and also by the bi-linear ansatz method \cite{PraCis2019};
for a mKdV nonlinearity, $p=2$, 
only approximate solutions have been found \cite{CisPraVilCar2018}.
No work has been done to-date on higher nonlinearities. 

The primary goal of the present paper is to develop and apply a new general method 
that yields numerous exact families of solitary wave solutions \eqref{inv.soln}. 
In total, 
nine solution families are obtained for $p=1$, 
eight families for $p=2$, 
four families for $p=3$, 
and a single family for $p=4$. 
All of the solutions are non-singular. 
The gKdV component $U(\z)$ has
\begin{itemize}
\item
sech-squared profile 
\item
bright/dark peak with exponential decay 
\item
width $w$ and height/depth $h$ related by $w^2 h\propto |\a|^{\frac{1}{p}}|\b|^{\frac{1}{p}-2}|\d|^{3-\frac{2}{p}}/|\e|$
\item
zero or non-zero (asymptotic) background, $b$
\end{itemize}
More interestingly, 
the magnitude of the LS component $|\Psi(\z)|$ has
\begin{itemize}
\item
bright/dark central peak
\item
asymptotic tail which is bright/dark relative to the central peak
\item 
in some cases, symmetric pairs of side peaks (up to $p$ in total) 
\end{itemize}
while the LS phase $\arg(\Psi(\z))$ has
\begin{itemize}
\item
a linear part $\frac{c}{2}\z$
\item
in some cases, a nonlinear part with a kink profile
\end{itemize}
For $p=2,3,4$, the solutions are new. 
For $p=1$, five families are new,
while the other four families contain the previously known solutions 
in \Ref{NisHojMimIke,CisPel2017} as particular subfamilies. 
Qualitatively novel features are the kink part in the LS phase 
and the presence of multiple symmetric side peaks in the LS amplitude. 

The new method consists of the following main steps: \\
(1) express the form of the travelling waves as a symmetry-invariant solution 
and obtain first integrals for the ODE system by use of multi-reduction symmetry theory
\cite{AncGan};
\\
(2) change to polar variables and apply a hodograph transformation that 
leads to a triangular (decoupled) system in which the base ODE has a polynomial form;
\\
(3) employ a power-balance technique to determine the general form for polynomial solutions of the base ODE; 
\\
(4) characterize conditions under which such solutions yield solitary waves;
\\
(5) with those conditions, solve an overdetermined algebraic system 
for the coefficients in the polynomial solutions 
along with the power $p$ 
and the parameters $\a,\b,\c,\d,\e$ in the original gKdV-LS system 
in addition to $c,\omega,b$ in the form of the travelling wave.

This algebraic system is nonlinear, with $p$ appearing in exponents, 
and so it is very non-trivial to solve. 
All solutions are found by employing the symbolic computer algebra package Crack \cite{Wol}
which is able to do non-polynomial case distinctions and splittings systematically. 
There is a further non-trivial aspect, stemming 
from the conditions for a solution to yield a solitary wave. 
Some of these conditions consist of inequalities on the parameters in the solution,
while each solution has a different subset of the parameters that are free, 
up to reparameterization freedom. 
An optimal choice of free parameters must be found so that 
all of the inequalities can be solved explicitly, 
thereby yielding the allowed kinematic region in parameter space. 
This is carried out by use of Maple,
which is also used to verify the final simplified explicit form 
for each family of solitary wave solutions. 

Some additional results are obtained. 
All Lie point symmetries and low-order conservation laws are derived for 
the coupled gKdV-LS system \eqref{u.eqn.phys}--\eqref{psi.eqn.phys}. 
Through these conservation laws, 
a variational principle for phase-modulated travelling waves \eqref{inv.soln} is found
and conserved quantities describing mass, charge, energy, and momentum
are evaluated for each family of solitary waves. 
The variational principle provides a starting point for analysis of stability of the solitary waves. 

The present method is applicable more generally to 
other coupled nonlinear dispersive wave equations 
as well as to nonlinear ODE systems of generalized H\'enon-Heiles form. 

The rest of the paper is organized as follows. 
Section~\ref{sec:prelim} discusses the Hamiltonian and Lagrangian formulation of 
the coupled gKdV-LS system \eqref{u.eqn.phys}--\eqref{psi.eqn.phys},
and gives the dimensionless form of this system which is used for subsequent results. 
Section~\ref{sec:symms.conslaws} gives the derivation of the Lie point symmetries and low-order conservation laws 
and describes the physical meaning of the conserved quantities. 
Section~\ref{sec:ODEsystem} states the variational principle for phase-modulated travelling waves 
and sets up the first two steps in the new method for finding solitary wave solutions. 
The remaining three steps in this method are carried out in section~\ref{sec:classifysolns},
and a classification of all of the solution families is provided. 
Section~\ref{sec:features} outlines the main features of each family,
including a few remarks on stability for these solitary wave solutions. 
Finally, some concluding remarks are made in section~\ref{sec:conclude}. 

Computational steps of the method are given in Appendix~\ref{sec:computation};
the output for each case $p=1,2,3,4$ is shown in Appendix~\ref{sec:output},
which also contains the solved inequalities giving the allowed kinematic region in parameter space for each solution family.

\section{Hamiltonian structure and dimensionless form}\label{sec:prelim}

The gKdV-LS system \eqref{u.eqn.phys}--\eqref{psi.eqn.phys} 
has a Hamiltonian structure
\begin{equation}\label{hamil.eqn}
\frac{\partial}{\partial t}\begin{pmatrix}
u \\ \psi
\end{pmatrix}
= \mathcal{D}\begin{pmatrix}
\delta H/\delta u \\ \delta H/\delta\bar\psi
\end{pmatrix}
\end{equation}
in which 
\begin{equation}\label{hamil.op}
\mathcal{D}={\rm diag}(D_x, -\tfrac{\sigma}{\gamma}i)
\end{equation} 
is a Hamiltonian operator,
and 
\begin{equation}
H=\int_\Rnum \big( \tfrac{1}{2}\b u_x^2 - \tfrac{1}{(p+1)(p+2)}\a u^{p+2} + (\d\c/\e) |\psi_x|^2 +\c u|\psi|^2 \big)\, dx
\end{equation}
is the Hamiltonian functional. 
The associated Poisson bracket is given by 
\begin{equation}
\{F,G\} = \Re \int_\Rnum 
\begin{pmatrix}\delta F/\delta u \\ \delta F/\delta\bar\psi \end{pmatrix}^\t \mathcal{D} \begin{pmatrix}\delta G/\delta u \\ \delta G/\delta\bar\psi \end{pmatrix}\,dx , 
\end{equation}
where $F$ and $G$ are arbitrary functionals of $u$, $\psi$, $\bar\psi$ and their $x$-derivatives. 
This bracket is skew and satisfies the Jacobi identity 
as a consequence of $\mathcal{D}$ being a Hamiltonian operator,
as follows from general results in the theory of Hamiltonian PDEs \cite{Olv-book}. 

A Lagrangian formulation can be obtained via introduction of a potential by $u=\v_x$. 
The system \eqref{u.eqn.phys}--\eqref{psi.eqn.phys} then takes the form 
\begin{align}
\v_{tx} + \a \v_x^p \v_{xx} + \b \v_{xxxx} & = \c (|\psi|^2)_x, 
\label{v.poteqn}
\\
i\psi_t + \d \psi_{xx} & = \e \v_x \psi, 
\label{psi.poteqn}
\end{align}
which is readily seen to be the Euler-Lagrange equations of the Lagrangian 
\begin{equation}
L = -\tfrac{1}{2}\v_t \v_x - \tfrac{1}{(p+1)(p+2)} \v_x^{p+2} +\tfrac{1}{2}\b \v_{xx}^2 
+\tfrac{1}{2}(\gamma/\sigma)i(\bar\psi_t \psi -\bar\psi\psi_t) +(\c \d/\e) |\psi_x|^2 
+ \c \v_x |\psi|^2  .
\end{equation}
Specifically, 
$\delta L/\delta \v = \v_{tx} + \a \v_x^p \v_{xx} + \b \v_{xxxx} -\c (|\psi|^2)_x$
and $i(\sigma/\gamma) \delta L/\delta \bar\psi = i\psi_t + \d \psi_{xx} -\e \v_x \psi$.

\subsection{Dimensionless variables}

It will be convenient to work in dimensionless variables,
whereby the number of constants in the system \eqref{u.eqn.phys}--\eqref{psi.eqn.phys}
is reduced from five to one. 

\begin{prop}
The dimensionless form of the gKdV-LS system \eqref{u.eqn.phys}--\eqref{psi.eqn.phys}
is given by 
\begin{align}
& u_t + s_1 u^p u_x + u_{xxx} + s_2 (|\psi|^2)_x =0 , 
\label{u.eqn}
\\
& i\psi_t + \psi_{xx} + k u \psi =0 , 
\label{psi.eqn}
\end{align}
where 
\begin{equation}\label{s1.s2}
s_1 =\begin{cases} 
1, & p=\text{ odd }\\ 
\sgn(\a)\sgn(\b), & p=\text{ even }\end{cases},
\quad
s_2 = \begin{cases} 
-\sgn(\a)\sgn(\c), & p=\text{ odd }\\ 
1, & p=\text{ even }\end{cases}
\end{equation}
and 
\begin{equation}\label{k}
k=
%\begin{cases}
%-\sgn(\a)\sgn(\b)\sgn(\d) \e |\b|^{2 -1/p} |\a|^{-1/p}/|\d|^{3 -2/p}, & p=\text{ odd }\\ 
%\sgn(\b)\sgn(\c)\sgn(\d) \e |\b|^{2 -1/p} |\a|^{-1/p}/|\d|^{3 -2/p}, & p=\text{ even }\\ 
%\end{cases} 
s_2\; \sgn(\b)\sgn(\c)\sgn(\d) \e |\beta|^{2 -1/p} |\alpha|^{-1/p}/|\kappa|^{3 -2/p} . 
\end{equation}
The transformation to this form is given by 
\begin{equation}
\begin{gathered}
\lambda_1=\b/\d,
\quad
\lambda_2=\b^2/\d^3,
\quad
|\lambda_3|=(\d^2/|\a\b|)^{1/p},
\quad
\lambda_4=\sqrt{(\d^2/|\b|)^{1+1/p}/(|\a|^{1/p}|\c|)},
\\
\sgn(\lambda_3)=\begin{cases}
\sgn(\a)\sgn(\b),& p \text{ is odd}
\\
-\sgn(\b)\sgn(\c),& p \text{ is even} 
\end{cases} .
\end{gathered}
\end{equation}
\end{prop}

Note that the sign $s_1$ of the nonlinear convective term in equation \eqref{u.eqn}
has been chosen to match the conventional form of the gKdV equation. 
The signs $s_1$ and $s_2$, along with $\sgn(k)$, control the sign properties of the Hamiltonian, 
as discussed in subsection~\ref{sec:conslaws}.

\begin{proof}
Consider the general transformation group 
\begin{equation}\label{transform.vars}
x\to \lambda_1 x,
\quad
t\to \lambda_2 t,
\quad
u\to \lambda_3 u,
\quad
\psi\to \lambda_4 \psi
\end{equation}
where each $\lambda$ is a non-zero parameter. 
The constants in the equations \eqref{u.eqn.phys}--\eqref{psi.eqn.phys}
transform as 
\begin{equation}\begin{gathered}
\a\to \lambda_1^{-1}\lambda_2^{\mathstrut}\lambda_3^p \a =\tilde\a, 
\quad
\b\to \lambda_1^{-3}\lambda_2^{\mathstrut} \b =\tilde\b, 
\\
\c\to  \lambda_1^{-1} \lambda_2^{\mathstrut} \lambda_3^{-1}\lambda_4^2 \c=\tilde\c,
\quad
\d\to  \lambda_1^{-2} \lambda_2^{\mathstrut}  \d=\tilde\d, 
\quad
\e\to \lambda_2 \lambda_3 \e=\tilde\e .
\end{gathered}
\end{equation}
First, scalings, with $\lambda_i>0$, 
can be used to put 
\begin{equation}
|\tilde\a| = |\tilde\b| = |\tilde\c| = |\tilde\d| = 1
\end{equation}
via 
\begin{equation}
\lambda_1=|\b|/|\d|,
\quad
\lambda_2=\b^2/|\d|^3,
\quad
\lambda_3=(\d^2/|\a\b|)^{1/p},
\quad
\lambda_4=\sqrt{(\d^2/|\b|)^{1+1/p}/(|\a|^{1/p}|\c|)}.
\end{equation}
Next, reflections, with $\lambda_i=\pm 1$, 
can be used to obtain 
\begin{equation}
\tilde\a =1 \text{ if $p$ is odd}, 
\quad
\tilde\b =1, 
\quad
\tilde\c =1 \text{ if $p$ is even}, 
\quad
\tilde\d =1
\end{equation}
via
\begin{equation}
\lambda_1=\sgn(\b)\sgn(\d),
\quad
\lambda_2=\sgn(\d),
\quad
\lambda_3=\begin{cases}
\sgn(\a)\sgn(\b),& p \text{ is odd}
\\
-\sgn(\b)\sgn(\c),& p \text{ is even}
\end{cases}.
\end{equation}
Under these transformations, 
$\tilde\e = \sgn(\d)\e|\a|^{-1/p}|\b|^{2-1/p}|\d|^{-3+2/p}\sgn(\lambda_3)$. 
\end{proof}

In \Ref{CisPel2017,PraCis2019}, which considers only the KdV nonlinearity, 
the coefficients are chosen to be 
$\a=-6$, $\b=1$, $\c=-1$, $\d=24/V_s$, $\e=24/V_s$,
where $V_s$ is an arbitrary positive constant. 
This corresponds to the values $s_1=1$, $s_2=-1$, $k=1/6$
for the scaled system \eqref{u.eqn}--\eqref{psi.eqn} with $p=1$.

\section{Symmetries and conservation laws}\label{sec:symms.conslaws}

The symmetry group of the scaled gKdV-LS system \eqref{u.eqn}--\eqref{psi.eqn}
will provide a foundation for the subsequent analysis. 
Generators of continuous point symmetries can be obtained by 
the standard Lie symmetry method \cite{Olv-book,BCA-book}, 
which yields the following result. 

\begin{prop}\label{prop:pointsymms}
(i) 
The continuous symmetries of the gKdV-LS system \eqref{u.eqn}--\eqref{psi.eqn}
for general $p\neq 0$ are generated by 
\begin{equation}\label{time.space.phase.symm}
\X_1=\partial_t, 
\quad
\X_2=\partial_x,
\quad
\X_3=i\psi\partial_\psi -i\bar\psi\partial_{\bar\psi}.
\end{equation}
Respectively, they comprise time-translations, space-translations, and phase rotations,
\begin{equation}\label{time.space.phase.symmgroup}
t\to t+\epsilon_1,
\quad
x\to x+\epsilon_2,
\quad
\psi\to \exp(i\epsilon_3)\psi,
\end{equation}
where $\epsilon_1$, $\epsilon_2$, $\epsilon_3$ $\in \Rnum$ are group parameters. 
(ii) 
Additional continuous symmetries arise only for $p=1$. 
They are generated by 
\begin{equation}\label{galilean.symm}
\X_4=t\partial_x + \partial_u + (s_1 k t+\tfrac{1}{2}x)(i\psi\partial_\psi -i\bar\psi\partial_{\bar\psi}),
\end{equation}
which constitutes a Galilean boost
\begin{equation}\label{galilean.symmgroup}
\quad
x\to x+\epsilon_4 t,
\quad
u\to u+\epsilon_4,
\quad
\psi\to \exp\big(i(\epsilon_4(kt+\tfrac{1}{2}x) +\epsilon_4{}^2\tfrac{1}{4}t)\big)\psi
\end{equation}
with group parameter $\epsilon_4$ $\in \Rnum$. 
\end{prop}

These transformations \eqref{time.space.phase.symmgroup} and \eqref{galilean.symmgroup}
generate, respectively, 
a three-dimensional abelian symmetry group %$\Rnum^2\times U(1)$ 
and a four-dimensional non-abelian symmetry group with the commutator structure
\begin{equation}
[\X_1,\X_4] = \X_2 + s_1 k \X_3,
\quad
[\X_2,\X_4] = \X_3 .
\end{equation}

The gKdV-LS system \eqref{u.eqn}--\eqref{psi.eqn} also possesses 
a discrete point symmetry $t\to -t$, $x\to -x$, $\psi\to \bar\psi$, 
which is a space-time reflection combined with complex conjugation.

\subsection{Conservation laws}\label{sec:conslaws}

A conservation law is a continuity equation 
\begin{equation}\label{conslaw}
D_t T + D_x X =0
\end{equation}
that holds for all formal solutions $(u(t,x),\psi(t,x))$ of the gKdV-LS system \eqref{u.eqn}--\eqref{psi.eqn},
where $D_t$ and $D_x$ denote total derivatives with respect to $t$ and $x$,
which act by the chain rule \cite{Olv-book,BCA-book}. 
Here $T$ is the conserved density and $X$ is the spatial flux; the pair $(T,X)$ is called the conserved current. 
In general, $T$ and $X$ can be functions of $t$, $x$, $u$, $\psi$, $\bar\psi$, and $x$-derivatives of $u$, $\psi$, $\bar\psi$,
with $t$-derivatives being eliminated through the equations \eqref{u.eqn} and \eqref{psi.eqn}. 

Conservation laws in which $T=D_x\Theta$ and $X=-D_t\Theta$ hold
for all formal solutions $(u(t,x),\psi(t,x))$ are trivial, 
since the continuity equation reduces to an identity via commutativity of derivatives. 
Trivial conservation laws thus contain no information about solutions. 
Any conservation laws that differ just by a trivial conservation law
can be viewed as being locally equivalent. 
Consequently, 
only non-trivial conservation laws up to local equivalence are of interest. 

Any non-trivial conservation law yields a corresponding conserved integral 
$\int_\Omega T\,dx$ which satisfies 
$\frac{d}{dt}\int_\Omega T\,dx = -X|_{\partial\Omega}$,
on any given spatial domain $\Omega\subset\Rnum$ of interest. 
Under appropriate boundary conditions posed at $\partial\Omega$ for solutions $(u(t,x),\psi(t,x))$, 
the conserved integral will be time-independent. 
In the situation when $\Omega=\Rnum$ is the real line, 
the necessary boundary conditions take the form of spatial decay conditions on 
$u(t,x)$ and $|\psi(t,x)|$ as $|x|\to\infty$, so that the spatial flux $X\to0$, whereby 
\begin{equation}
\frac{d}{dt}\int_\Rnum T\,dx = 0 .
\end{equation}
Note that the conserved integral $\int_\Rnum T\,dx$ is unchanged 
if $T$ is replaced by a locally equivalent conserved density,
satisfying suitable spatial decay conditions. 
The differential order of a conserved integral will refer to 
the minimum of the differential orders of all conserved densities 
that are locally equivalent to $T$. 
Conserved integrals of most physical interest typically have a differential order 
that is no higher than the differential order of the Hamiltonian. 

For the method applied in subsequent sections for deriving 
explicit solutions $(u(t,x),\psi(t,x))$ of the gKdV-LS system \eqref{u.eqn}--\eqref{psi.eqn}, 
it will be essential to know both the conserved density $T$ and the spatial flux $X$
in a conservation law. 

The generators of continuous point symmetries can be used to obtain 
conservation laws of the gKdV-LS system \eqref{u.eqn}--\eqref{psi.eqn} 
through the Hamiltonian version of Noether's theorem. 
Additional conservation laws arise from the Casimirs of the Hamiltonian structure. 

Alternatively, conservation laws can be obtained directly by use of the standard multiplier method \cite{Olv-book,BCA-book,AncBlu2002,Anc-review,Wol}, 
which does not rely on existence of any variational structure. 

Both of these methods for finding conservation laws yield the following result. 

\begin{prop}\label{prop:conslaws}
For general $p\neq0$, 
all non-trivial conservation laws of the gKdV-LS system \eqref{u.eqn}--\eqref{psi.eqn} 
with conserved densities of differential order at most $1$
are spanned by the conserved currents
\begin{align}
& \begin{aligned}
T_1 = u, 
\quad
X_1 = u_{xx} + s_1\tfrac{1}{p+1} u^{p+1} + s_2|\psi|^2 ;
\end{aligned}
\label{T.X.mass}
\\
& \begin{aligned}
T_2 = \tfrac{1}{2} |\psi|^2, 
\quad
X_2 = |\psi|^2\arg(\psi)_x ;
\end{aligned}
\label{T.X.charge}
\\
& \begin{aligned}
T_3 & = \tfrac{1}{2} k  u^2 - s_2 |\psi|^2\arg(\psi)_x , 
\\
X_3 & = k( uu_{xx} -\tfrac{1}{2} u_x^2 + s_1 \tfrac{1}{p+2} u^{p+2} ) - s_2 |\psi_x|^2 + s_2 |\psi|^2 \arg(\psi)_t ;
\end{aligned}
\label{T.X.mom}
\\
& \begin{aligned}
T_4 & = \tfrac{1}{2} u_x^2 - s_1\tfrac{1}{(p+1)(p+2)} u^{p+2} +(s_2/k) |\psi_x|^2 -s_2 u|\psi|^2 ,
\\
X_4 & =  -\tfrac{1}{2}(u_{xx} + s_1 \tfrac{1}{p+1} u^{p+1}+ s_2 |\psi|^2)^2
- u_t u_x -(s_2/k) (\psi_t\bar\psi_x +\psi_x\bar\psi_t) .
\end{aligned}
\label{T.X.ener}
\end{align}
Additional conservation laws of differential order at most $1$ 
arise only for $p=1$. 
They are spanned by 
\begin{equation}
\begin{aligned}
T_5 & = s_2k( \tfrac{1}{2} t u^2 - s_1 x u )
-t |\psi|^2\arg(\psi)_x +(s_1k t + \tfrac{1}{2} x) |\psi|^2 ,
\\
X_5 & =  s_2 k( (t u -s_1 x) u_{xx} +s_1 u_x -\tfrac{1}{2} t u_x^2 -\tfrac{1}{2} x u^2 +s_1 \tfrac{1}{3} t u^3 ) 
-t |\psi_x|^2 -s_1 kx |\psi|^2 
\\&\quad
%-\tfrac{1}{2} i t (\psi_t\bar\psi -\bar\psi_t\psi) 
+t |\psi|^2\arg(\psi)_t 
%-i(s_1 kt +\tfrac{1}{2} x) (\psi_x\bar\psi -\bar\psi_x\psi) 
+(s_1 2 kt + x) |\psi|^2 \arg(\psi)_x . 
\end{aligned}
\label{T.X.galmom}
\end{equation}
The corresponding conserved integrals consist of 
\begin{align}
& {M}=\int_\Rnum u\, dx\quad\text{mass,}
\label{mass}
\\
& {J}=\int_\Rnum \tfrac{1}{2} |\psi|^2 \, dx\quad\text{charge,}
\label{charge}
\\
& {P}=\int_\Rnum \big(\tfrac{1}{2} u^2 -s_2\tfrac{1}{k} |\psi|^2\arg(\psi)_x\big)\, dx\quad\text{momentum,}
\label{momentum}
\\
& {H}=\int_\Rnum \big(\tfrac{1}{2} u_x^2 - s_1\tfrac{1}{(p+1)(p+2)} u^{p+2}  +s_2\tfrac{1}{k} |\psi_x|^2 -s_2 u|\psi|^2\big)\, dx \quad\text{(Hamiltonian) energy,}
\label{energy}
\end{align}
for $p\neq0$, 
and 
\begin{equation}
\begin{aligned}
{G}=\int_\Rnum \big(s_2k( \tfrac{1}{2} t u^2 - s_1 x u ) -t|\psi|^2\arg(\psi)_x 
+(s_1k t + \tfrac{1}{2} x) |\psi|^2\big)\, dx \quad\text{Galilean momentum}
\end{aligned}
\end{equation}
for $p=1$. 
\end{prop}

With respect to the Hamiltonian structure \eqref{hamil.eqn}--\eqref{hamil.op},
mass is a Casimir, 
while charge, momentum, energy, and Galilean momentum 
arise respectively from the 
phase rotation symmetry, space-translation symmetry, time-translation symmetry.
and Galilean boost symmetry. 
Two remarks on these conserved integrals will be worthwhile.

The energy (Hamiltonian) will be non-negative when $s_1=-1$, $s_2=\sgn(k)=-\sgn(u)$,
and otherwise it will have an indefinite sign. 
The sign of $s_1$ controls the gKdV portion of the energy, 
while the signs of $s_2$ and $k$ control the LS portion. 
When the sign is indefinite,
conservation of energy does not preclude blow-up of $|u|$, $|u_x|$, $|\psi|$, $|\psi_x|$ 
caused by concentration of the energy density, which is referred to as \emph{focusing};
in contrast, when the sign is non-negative,
conservation of energy prevents such blow-up, which is referred to as \emph{defocusing}
(see e.g. \Ref{Tao-book}).

The Galilean momentum can be expressed as 
${G}=kt( s_2 {P} +s_1 {J}) -\chi(t)$
where $\chi(t) = \int_\Rnum (s_1 s_2k u - \tfrac{1}{2} |\psi|^2)x\,dx$
is like a center of mass-charge. 
In particular, 
under decay conditions such that ${G}$, ${P}$, ${J}$ are conserved integrals, 
$\chi(t)$ satisfies $\frac{d}{dt}\chi(t) = k( s_2 {P} +s_1 {J})$.
This implies $\chi(t)=\chi(0) + t V$, where $V= k( s_2 {P} +s_1 {J})$ is the speed
at which the center of mass-charge moves.

\section{ODE system for phase-modulated travelling waves}\label{sec:ODEsystem}

Phase-modulated travelling waves \eqref{inv.soln}
have a symmetry characterization that they are invariant under 
a Galilean transformation $x\to x+c\epsilon$, $t\to t+\epsilon$ 
to a reference frame moving with speed $c$, 
combined with a phase rotation $\psi\to e^{-i\w\epsilon}\psi$
having angular speed $\w$. 
Specifically, the symmetry generator
\begin{equation}\label{symm}
\X=\partial_t + c\partial_x -i\w\psi\partial_\psi +i\w\bar\psi\partial_{\bar\psi}
\end{equation}
has the invariants $U=u$, $\Psi=e^{i\w t}\psi$, $\z=x-ct$,
and hence a phase-modulated travelling wave is an invariant solution of 
the gKdV-LS system \eqref{u.eqn}--\eqref{psi.eqn} 
under this symmetry. 

Substitution of the invariant form \eqref{inv.soln} into the coupled equations \eqref{u.eqn} and \eqref{psi.eqn} 
yields a coupled nonlinear ODE system for $U(\z)$ and $\Psi(\z)$:
\begin{subequations}\label{U.Psi.sys}
\begin{align}
& U'''+ (s_1 U^p-c)U' + s_2(|\Psi|^2)' =0 , 
\\
& \Psi'' -ic\Psi'+(kU +\w)\Psi =0 .
\end{align}
\end{subequations}
It will be useful to consider the polar form of this system, 
which arises from expressing $\Psi$ in amplitude-phase form 
\begin{equation}\label{amplitudephase}
\Psi=\H \exp(i\Y) .
\end{equation}
The polar variables $\H(\z)$ and $\Y(\z)$, along with $\V(\z)$, satisfy 
the coupled nonlinear ODEs
\begin{subequations}\label{U.A.Phi.sys}
\begin{align}
& \V'''+ (s_1 \V^p-c)\V' + 2s_2 \H \H' =0 , 
\label{U.ode}
\\
& \H'' +(k\V +c\Y' -\Y'{}^2 +\w)\H =0 , 
\label{A.ode}
\\
& \H\Y'' + (2\Y'-c)\H'=0 . 
\label{Phi.ode}
\end{align}
\end{subequations}

\subsection{Variational formulation and symmetries}

The ODE system \eqref{U.A.Phi.sys} for phase-modulated travelling waves 
has a variational formulation in terms of the conserved integrals
\eqref{mass}--\eqref{energy}
for mass, charge, momentum, and energy
evaluated for solutions $\V(\z)$, $\H(\z)$, $\Y(\z)$. 
These conserved integrals consist of
\begin{align}
& {M}[\V] =\int_{-\infty}^{\infty} \V\, d\z, 
\label{M}\\
& {J}[\H] =\int_{-\infty}^{\infty}\tfrac{1}{2} \H^2 \, d\z, 
\label{J}\\
& {P}[\V,\H,\Y] =\int_{-\infty}^{\infty}( \tfrac{1}{2} \V^2 -s_2\tfrac{1}{k} \H^2\Y' )\, d\z, 
\label{P}\\
& {H}[\V,\H,\Y] =\int_{-\infty}^{\infty}( \tfrac{1}{2} \V'{}^2 - s_1\tfrac{1}{(p+1)(p+2)} \V^{p+2}  +s_2\tfrac{1}{k} (\H'{}^2 +\H^2 \Y'{}^2) -s_2 \V\H^2 )\, d\z .
\label{H}
\end{align}

Consider the conserved integral 
\begin{equation}\label{W}
W[\V,\H,\Y] = {H}[\V,\H,\Y] + c {P}[\V,\H,\Y] -s_2 (2\omega/k) {J}[\H] + C_1 {M}[\V] ,
\end{equation}
where $C_1$ is an arbitrary constant. 
Then the three ODEs \eqref{U.ode}, \eqref{A.ode}, \eqref{Phi.ode} are 
respectively equivalent to the Euler-Lagrange equations 
\begin{equation}\label{W.ELeqns}
\begin{pmatrix}
\delta W/\delta\V 
\\
\delta W/\delta\H
\\
\delta W/\delta\Y
\end{pmatrix}
=0 .
\end{equation}
In particular, the first of these equations yields the integral of ODE \eqref{U.ode} 
with $C_1$ being the constant of integration. 

This variational formulation is a generalization of the known variational formulations
respectively for travelling waves of the gKdV equation 
\cite{BonSouStr}
and phase-modulated travelling waves of the Schr\"odinger equation
\cite{CazLio}. 

The Euler-Lagrange equations \eqref{W.ELeqns} manifestly inherit 
a translation symmetry and a phase-shift symmetry 
\begin{equation}\label{ODE.symms}
\X_{\rm trans.} = \partial_\z,
\quad
\X_{\rm phas.} = \partial_\Y 
\end{equation}
from the symmetries of the gKdV-LS system \eqref{u.eqn}--\eqref{psi.eqn}. 
A complete classification of all continuous symmetries is straightforwardly obtained
by the Lie symmetry method. 

\begin{prop}\label{prop:ODE.pointsymms}
(i) 
For general $p\neq 0$, 
the continuous symmetries of the Euler-Lagrange equations \eqref{W.ELeqns} 
are generated by the translation and phase-shift symmetries \eqref{ODE.symms},
which yield the transformation groups
$\z \to \z +\epsilon$ and $\Y \to \Y +\epsilon$, 
where $\epsilon\in\Rnum$ is the group parameter for each transformation. 
(ii) 
Additional symmetries exist only when $p=1$.
They are generated by 
\begin{equation}\label{ODE.symms.pis1}
\begin{aligned}
& \X_{\rm scal.} = -\z\partial_\z + 2(\V - s_1 c)\partial_\V + 2\H\partial_\H -\frac{c}{2}\z\partial_\Y,
\\
& 
\omega = -s_1 kc -\frac{c^2}{4},
\quad
C_1 = -s_1 \frac{c^2}{2},
\quad
p=1 , 
\end{aligned}
\end{equation}
which yields the scaling-type transformation group
$\z\to \lambda^{-1}\z$, 
$\V\to \lambda^{2}\V +s_1 c(1-\lambda^{2})$, 
$\H\to \lambda^{2}\H$, 
$\Y\to \Y +(1-\lambda^{-1})c\z/2$, 
where $\lambda\neq0\in\Rnum$ is the group parameter. 
\end{prop}

These symmetries have a key role in the analysis of stability of solutions.

\subsection{Reduction to a decoupled second-order ODE}

Ideally, the goal would be to solve the ODE system \eqref{U.A.Phi.sys} 
in explicit form for $\V(\z)$, $\H(\z)$, $\Y(\z)$. 
Note that physically non-trivial solutions must have 
$\V'\not\equiv0$ and $\H'\not\equiv0$. 

This system consists of one third-order ODE \eqref{U.ode} and two second-order ODEs \eqref{A.ode} and \eqref{Phi.ode}. 
Therefore, finding the general solution would require a total of seven integrations. 
As noted in the work \Ref{CisPel2017,CisPraVilCar2018}, 
inspection of the ODEs suggest that only a few explicit integrations are possible. 

Nevertheless, remarkably, 
the system \eqref{U.A.Phi.sys} can be reduced to a single second-order ODE 
from which $\V(\z)$, $\H(\z)$, $\Y(\z)$ can be obtained by quadratures. 
This reduction is found by the following new method which involves two main steps. 

In the first step, symmetry multi-reduction is applied, 
which gives a set of three functionally independent first integrals. 
These first integrals arise from conservation laws of 
the gKdV-LS system \eqref{u.eqn}--\eqref{psi.eqn} 
such that the conservation law is invariant (up to local equivalence) 
under the symmetry \eqref{symm} which characterizes 
phase-modulated travelling waves. 
The multi-reduction method yields the first integrals directly, 
without the need to know in advance any conservation laws. 
From the first integrals, 
an explicit reduction to a system of one second-order ODE and two first-order ODEs 
is obtained. 

The second step starts from observing that 
the resulting second-order ODE in the first step 
can be expressed in the form of a nonlinear oscillator equation. 
This equation admits an energy-type first integral, 
which can be combined with the remaining two first-order ODEs to get 
a decoupled second-order ODE by employing a hodograph transformation.  
The resulting system of ODEs is triangular such that every solution of 
the transformed second-order ODE yields a solution $(\V(\z)$, $\H(\z)$, $\Y(\z))$
by quadratures. 
Furthermore, 
two of these quadratures can be carried out explicitly to obtain $\V(\z)$ and $\H(\z)$, 
while $\Y(\z)$ is expressed as a final quadrature in terms of $(\V(\z),\H(\z))$. 

The preceding steps will be carried out in the next subsections.

\subsection{First integrals from symmetry multi-reduction}

The first step consists of deriving first integrals of the ODE system \eqref{U.A.Phi.sys}
by use of the method of symmetry multi-reduction \cite{AncGan}. 
This method relies on the property if a conservation law \eqref{conslaw}
of the gKdV-LS system \eqref{u.eqn}--\eqref{psi.eqn} 
is invariant under the symmetry \eqref{symm} that characterizes the ODE system, 
then the conserved density $T$ and the flux $X$ yield a first integral 
$\tfrac{dC}{d\xi}=0$
given by 
\begin{equation}\label{FI.fromTX}
C=(X-cT)|_{u=U,\psi=e^{i\w t}\Psi}
= \const
\end{equation}
for all solutions $(\V(\z)$, $\H(\z)$, $\Y(\z))$. 
This expression \eqref{FI.fromTX} has the physical meaning of the net flux 
in the moving reference frame of travelling wave. 

All symmetry-invariant conservation laws can be derived in a systematic way 
based on the use of multipliers. 
Specifically, there is an explicit condition characterizing multipliers that yield symmetry-invariant conservation laws. 

For the gKdV-LS system \eqref{u.eqn}--\eqref{psi.eqn}, 
a multiplier is a pair of functions 
$(Q^u,Q^\psi)$ of $t$, $x$, $u$, $\psi$, $\bar\psi$, and $x$-derivatives of $u$, $\psi$, $\bar\psi$,
such that 
\begin{equation}
\begin{aligned}
& (u_t + s_1 u^p u_x + u_{xxx} + s_2 (|\psi|^2)_x)Q^u
+ (i\psi_t + \psi_{xx} + k u \psi)Q^\psi
+ (-i\bar\psi_t + \bar\psi_{xx} + k u \bar\psi)\bar Q^\psi
\\&\quad
= D_t T + D_x X
\end{aligned}
\end{equation}
for some functions $T$ and $X$. 
Since the system \eqref{u.eqn}--\eqref{psi.eqn} is comprised of evolution equations, 
the general theory of multipliers shows that, firstly, 
every non-trivial conservation law (up to local equivalence) 
will arise from a multiplier,
and that, secondly, 
multipliers are related to $(T,X)$ by a variational derivative, 
$Q^u=E_u(T)$ and $Q^\psi = E_{\bar\psi}(T)$,
where $E_v$ denotes the Euler operator (variational derivative) with respect to a variable $v$. 
These expressions show that any conserved density of differential order at most $1$ 
will arise from a multiplier of differential order at most $2$. 
Hence, only multipliers given by functions of 
$t$, $x$, $u$, $\psi$, $\bar\psi$, and $x$-derivatives of $u$, $\psi$, $\bar\psi$ up to second-order 
will be considered hereafter. 

All multipliers of differential order at most $2$ are the solutions of the linear system of determining equations
\begin{equation}\label{multr.deteqns}
\begin{aligned}
& E_v\big( (u_t + s_1 u^p u_x + u_{xxx} + s_2 (|\psi|^2)_x)Q^u
+ (i\psi_t + \psi_{xx} + k u)Q^\psi + (-i\bar\psi_t + \bar\psi_{xx} + k u \bar\psi)\bar Q^\psi 
\big)
\\&
=0,
\quad
v=\{u,\psi,\bar\psi\} . 
% E_\psi\big( (u_t + s_1 u^p u_x + u_{xxx} + s_2 (|\psi|^2)_x)Q^u + (i\psi_t + \psi_{xx} + k u)Q^\psi + (-i\bar\psi_t + \bar\psi_{xx} + k u \bar\psi)\bar Q^\psi \big)=0
% E_{\bar\psi}\big( (u_t + s_1 u^p u_x + u_{xxx} + s_2 (|\psi|^2)_x)Q^u + (i\psi_t + \psi_{xx} + k u)Q^\psi + (-i\bar\psi_t + \bar\psi_{xx} + k u \bar\psi)\bar Q^\psi \big)=0
\end{aligned}
\end{equation}
These three equations split with respect to $u_t$, $\psi_t$, $\bar\psi_t$, and their $x$-derivatives, 
yielding an overdetermined linear system for $(Q^u,Q^\psi)$. 
For each solution, a corresponding conserved current $(T,X)$ can be recovered 
by use of a homotopy integral formula. 

The theory of symmetry multi-reduction shows that the condition
for a multiplier to yield a symmetry-invariant conservation law is given by 
$\pr\X(Q^u,Q^\psi) =0$,
where $\pr\X$ is the prolongation of the symmetry generator \eqref{symm}.
This condition is a linear system of two equations 
\begin{subequations}\label{multr.inveqns}
\begin{gather}
i\w\big( \psi_{xx}Q^u_{\psi_{xx}} -\bar\psi_{xx}Q^u_{\bar\psi_{xx}}
+ \psi_{x}Q^u_{\psi_{x}} -\bar\psi_{x}Q^u_{\bar\psi_{x}}
+ \psi Q^u_{\psi} -\bar\psi Q^u_{\bar\psi} \big) -c Q^u_x - Q^u_t =0, 
\\
i\w\big( \psi_{xx}Q^\psi_{\psi_{xx}} -\bar\psi_{xx}Q^\psi_{\bar\psi_{xx}}
+ \psi_{x}Q^\psi_{\psi_{x}} -\bar\psi_{x}Q^\psi_{\bar\psi_{x}}
+ \psi Q^\psi_{\psi} -\bar\psi Q^\psi_{\bar\psi} +Q^\psi \big) 
-c Q^\psi_x -Q^\psi_t =0 . 
\end{gather}
\end{subequations}
Hence, all symmetry-invariant conservation laws arise from the solutions of 
the overdetermined linear system given by splitting the five equations \eqref{multr.deteqns}--\eqref{multr.inveqns}
with respect to $u_t$, $\psi_t$, $\bar\psi_t$, and their $x$-derivatives. 
Each solution $(Q^u,Q^\psi)$ determines a symmetry-invariant conserved current $(T,X)$ 
which in turn yields a first integral \eqref{FI.fromTX} of the ODE system \eqref{U.A.Phi.sys}. 

The result is that the four conservation laws \eqref{T.X.mass}, \eqref{T.X.charge}, \eqref{T.X.mom}, \eqref{T.X.ener} 
from Proposition~\ref{prop:conslaws}
are symmetry invariant, 
while the fifth conservation law \eqref{T.X.galmom} which holds only for $p=1$ 
is not symmetry invariant. 
Hence, symmetry multi-reduction yields four first integrals. 
In polar variables, they are given by 
\begin{align}
& \begin{aligned}
C_1 = \V'' + s_2 \H^2 -c\V + s_1\tfrac{1}{p+1} \V^{p+1} , 
\end{aligned}
\\
& \begin{aligned}
C_2 = \H^2 (\Y' -\tfrac{1}{2} c) , 
\end{aligned}
\\
& \begin{aligned}
C_3 & = k( \V\V''  -\tfrac{1}{2} \V'{}^2 -\tfrac{1}{2} c \V^2 + s_1 \tfrac{1}{p+2} \V^{p+2} ) - s_2 \H'{}^2 - s_2 \H^2 (\Y'{}^2 +\w) , 
\end{aligned}
\\
& \begin{aligned}
C_4 & =  \tfrac{1}{2} k (\V'' + s_1 \tfrac{1}{p+1} \V^{p+1}+s_2\H^2)^2
-kc( \tfrac{1}{2} \V'{}^2 + s_1 \tfrac{1}{(p+1)(p+2)} \V^{p+2} )
\\&\qquad 
-s_2 c \H'{}^2 -s_2 \H^2 (c\Y'{}^2 +2\w \Y' +kc \V) . 
\end{aligned}
\end{align}
It is straightforward to see that these first integrals satisfy an algebraic relation 
\begin{equation}
\tfrac{1}{2} k C_1{}^2 -s_2 2\w C_2 + c C_3 -C_4=0 , 
\end{equation}
and so only three of them are functionally independent. 
This leads to the following result. 

\begin{lem}\label{lem:conslaws.reduction}
The conservation laws \eqref{T.X.mass}, \eqref{T.X.charge}, \eqref{T.X.mom}, \eqref{T.X.ener} 
of the gKdV-LS system \eqref{u.eqn}--\eqref{psi.eqn} 
yield three functionally independent first integrals of the ODE system \eqref{U.A.Phi.sys}
for phase-modulated travelling waves \eqref{inv.soln}. 
This gives a reduction of this system to one second-order ODE and two first-order ODEs
\begin{align}
\label{V''.eqn}
&\begin{aligned}
\V'' = c \V -s_1\tfrac{1}{p+1}\V^{p+1} -s_2 \H^2 +C_1 , 
\end{aligned}
\\
\label{V'.eqn}
& \begin{aligned}
\tfrac{1}{2} k \V'{}^2 +s_2 \H'{}^2 & = 
k C_1 \V + \tfrac{1}{2} k c \V^2 -s_1 \tfrac{1}{(p+2)(p+1)} k\V^{p+2}
-s_2 k \V \H^2 
\\&\qquad
- s_2 (\tfrac{1}{4} c^2+\w) \H^2 
-s_2 C_2{}^2 \H^{-2} -\tilde C_3 , 
\end{aligned}
\\
\label{Y'.eqn}
& \begin{aligned}
\Y' = \tfrac{1}{2} c +C_2 \H^{-2} , 
\end{aligned}
\end{align}
where $C_1$, $C_2$, $\tilde C_3$ ($=C_3+s_2 c C_2$) are free parameters. 
\end{lem}

This completes the first step in the analysis.

\subsection{Hodograph transformation to triangular form} 

The second step consists of showing that the ODE system \eqref{V''.eqn}--\eqref{Y'.eqn} 
can be decoupled into a triangular form, 
without any assumptions. 

Consider a hodograph transformation 
\begin{equation}\label{HV.rel}
\H=F(\V) . 
\end{equation}
Note that, since non-trivial solutions of the ODE system \eqref{U.A.Phi.sys}
must have $\H\neq\const$, 
this transformation thereby exists (at least) locally. 

Substitution of relation \eqref{HV.rel} into the system yields
\begin{align}
\label{V''.F.eqn}
&\begin{aligned}
\V'' = c \V -s_1\tfrac{1}{p+1}\V^{p+1} -s_2 F(\V)^2 +C_1 , 
\end{aligned}
\\
\label{V'.F.eqn}
& \begin{aligned}
\V'{}^2 & = 
\frac{ ( k C_1 \V + \tfrac{1}{2} k c \V^2 -s_1 \tfrac{1}{(p+2)(p+1)} k\V^{p+2}
- s_2 ( (\tfrac{1}{4} c^2+\w) + k \V )F(\V)^2 -\tilde C_3 )F(\V)^2 -s_2 C_2{}^2 }
{ F(\V)^2 (\tfrac{1}{2} k +s_2 F'(\V)^2) } , 
\end{aligned}
\\
\label{Y'.F.eqn}
& \begin{aligned}
\Y' = \tfrac{1}{2} c +C_2 F(\V)^{-2} . 
\end{aligned}
\end{align}
The second-order ODE \eqref{V''.F.eqn} now has the form of a nonlinear oscillator equation. 
It possesses the first integral 
\begin{equation}
\tilde C_4 = \tfrac{1}{2} \V'{}^2 + s_2 G(\V)  + s_1 \tfrac{1}{(p+1)(p+2)} \V^{p+2} -\tfrac{1}{2} c\V^2 -C_1 \V , 
\end{equation}
where $\tilde C_4$ is a constant, 
and where
\begin{equation}\label{GF.rel}
G(\V)=\int F(\V)^2 \, d\V . 
\end{equation}
This gives
\begin{equation}\label{V''.G.reduc}
\V'{}^2= - s_2 2 G(\V) -s_1 \tfrac{2}{(p+1)(p+2)} \V^{p+2} +c \V^2 + 2C_1 \V+2\tilde C_4 . 
\end{equation}
Equating equations \eqref{V''.G.reduc} and \eqref{V'.F.eqn}, 
and replacing 
\begin{equation}
F(\V)^2 =G'(\V),
\quad
F'(\V)=2G''(\V)/F(\V), 
\label{FG.rels}
\end{equation}
yields 
\begin{equation}\label{G.eqn}
\begin{aligned}
& \tfrac{1}{2}( s_2 G(\V) +s_1 \tfrac{1}{(p+1)(p+2)} \V^{p+2} -\tfrac{1}{2} c \V^2 - C_1 \V -\tilde C_4 )G''(\V)^2
\\&
=( k \V + \tfrac{1}{4} c^2+\w ) G'(\V)^2 - ( k G(\V) -s_2 (\tilde C_3 + k\tilde C_4) )G'(\V) + C_2{}^2 , 
\end{aligned}
\end{equation}
which is a second-order ODE for $G(\V)$. 
Hence, the following result has been obtained. 

\begin{lem}\label{lem:triangular.reduction}
The hodograph transformation 
\begin{equation}\label{H.G.rel}
\H = \sqrt{G'(\V)}
\end{equation}
converts the reduced ODE system \eqref{V''.eqn}--\eqref{Y'.eqn} 
into a triangular form consisting of 
a second-order ODE \eqref{G.eqn} for $G(\V)$, 
a first-order ODE \eqref{V''.G.reduc} for $\V(\z)$, 
and an elementary ODE \eqref{Y'.F.eqn} for $\Y(\z)$. 
\end{lem}

Lemma~\ref{lem:triangular.reduction} combined with Lemma~\ref{lem:conslaws.reduction}
effectively provides a reduction of the original ODE system \eqref{V''.eqn}--\eqref{Y'.eqn}
to a single second-order ODE \eqref{G.eqn}. 
As a consequence, 
phase-modulated travelling waves \eqref{inv.soln}
can be obtained by the following direct method. 

\begin{thm}\label{thm:reduction}
Any solution of the second-order ODE \eqref{G.eqn} for $G(\V)$ 
yields a solution of the ODE system \eqref{V''.eqn}--\eqref{Y'.eqn} 
by quadratures:
First, integrate ODE \eqref{V''.G.reduc}, which is separable, to get $\V(\z)$. 
Second, substitute $\V(\z)$ into transformation \eqref{H.G.rel} to get $\H(\z)$. 
Last, substitute $\H(\z)$ into ODE \eqref{Y'.eqn} and integrate to get $\Y(\z)$.
The resulting expressions $(\V(\z),\H(\z),\Y(\z))$ give 
a phase-modulated travelling wave 
\begin{equation}\label{soln}
u=U(\z),
\quad
\psi=e^{i(\w t +\Y(\z))} \H(\z),
\quad
\z=x-ct . 
\end{equation}
\end{thm}

Since non-trivial solutions of the ODE system \eqref{V''.eqn}--\eqref{Y'.eqn} 
must have $\H\neq\const$, 
corresponding solutions of the second-order ODE \eqref{G.eqn} for $G(\V)$ 
must satisfy 
\begin{equation}\label{G.cond}
G''(\V)\not\equiv0
\end{equation}

\section{Abundant solitary wave solutions for nonlinearity powers $p=1,2,3,4$}\label{sec:classifysolns}

The main result in Theorem~\ref{thm:reduction} reduces the problem of obtaining 
explicit phase-modulated travelling waves \eqref{soln}
to the much more tractable problem of finding explicit solutions of 
the decoupled second-order ODE \eqref{G.eqn}. 
Moreover, this ODE admits abundant polynomial solutions, 
which lead to explicit families of phase-modulated solitary waves,
on zero and non-zero backgrounds, in the cases $p=1,2,3,4$. 
By comparison, 
starting with an ansatz for solutions of the original ODE system \eqref{U.A.Phi.sys}, 
as seen in \Ref{CisPel2017,CisPraVilCar2018}, 
yields only a few solutions in the case $p=1$
and no solutions for $p\geq 2$. 

In the next two subsections, 
the general polynomial form for solutions will be introduced
and the types of phase-modulated solitary waves that arise from it 
will be characterized. 
In the final four subsections, 
the explicit solutions of the second-order ODE \eqref{G.eqn} will be summarized
in each of the cases $p=1,2,3,4$, 
which leads to the phase-modulated solitary waves 
given in section~\ref{sec:features}.

\subsection{Polynomial solutions of the decoupled ODE}\label{sec:ansatz}

The second-order ODE \eqref{G.eqn} for $G(\V)$ 
is quadratic in $G$, $G'$, $G''$, and also is polynomial in $\V$ when $p$ is a positive integer. All polynomial solutions for $G(\V)$ can be obtained by the following analysis. 

First, the possible degree of a polynomial solution 
is determined by balancing powers in the ODE \eqref{G.eqn}. 
Suppose that the highest power of $\V$ in $G(\V)$ is $\V^q$, 
with $q>1$ due to condition \eqref{G.cond}. 
Then, on the lefthand side in the ODE \eqref{G.eqn}, 
the terms containing highest powers of $\V$ are 
$G(\V)G''(\V)^2$ and $\V^{p+2}G''(\V)^2$, since $p\geq 1$. 
Their leading powers are, respectively, $3q-4$ and $2q+p-2$. 
On the righthand side, the terms containing highest powers of $\V$ are
$\V G'(\V)^2$ and $G(\V)G'(\V)$, 
both of which have the leading power $2q-1$. 
There are now several different possibilities for balancing powers. 

The first possibility is to balance the leading powers on opposite sides. 
If $2q+p-2>3q-4$,
then this balance is given by $2q+p-2=2q-1>3q-4$, 
which implies $p=1$ and $q< 3$. 
If $2q+p-2\leq 3q-4$,
then the balance is $3q-4=2q-1\geq 2q+p-2$
which implies $q=3$ and $p\leq 1$. 
Thus, a consistent balance holds only for $p=1$ and $q\leq 3$. 

The second possibility is that the two terms with leading powers $2q+p-2$ and $3q-4$
cancel on the lefthand side, whereby the term with the subleading power 
must be balanced by the leading power term on the righthand side. 
In this case, equality of the powers $2q+p-2=3q-4$ implies $q=p+2$. 
Suppose the subleading term in $G(\V)$ has power $l<q$, $l\geq 0$. 
Then the subleading term on the lefthand side will be 
the leading term in $(G(\V)+s_1s_2\tfrac{1}{(p+1)(p+2)}\V^{p+2})G''(\V)^2$, 
with corresponding highest power $l+2q-4=l+2p$. 
The leading terms on the righthand side have highest power $2q-1= 2p+3$. 
Hence, the balance gives $l=3$, with no restriction on $p$. 

The last possibility is that the two terms with leading power $2q-1$ 
on the righthand side cancel each other, 
whereby the term with the subleading power 
must be balanced by the leading power term on the lefthand side. 
In this case, the cancellation requires that the highest power term in 
$k\V G'(\V)^2 -k G(\V)G'(\V)= kG'(\V)(\V G'(\V)- G(\V))$ vanishes. 
Clearly, this implies that $q=1$, and thus no balance exists with $q>1$. 

The preceding possibilities can be merged into the form 
\begin{equation}\label{G.ansatz}
G(\V)=-s_1 s_2 \tfrac{1}{(p+1) (p+2)}\V^{p+2} +\tfrac{1}{3}g_3 \V^3 +\tfrac{1}{2}g_2 \V^2 +g_1 \V +g_0 , 
\end{equation}
where the coefficients $g_0, g_1, g_2, g_3$ are constant parameters to be determined. 
When this polynomial form is substituted in the ODE \eqref{G.eqn}, 
a polynomial equation of degree $2p+3$ in $\V$ is obtained. 
All solutions of interest will be listed after the discussion in the next subsection.

\subsection{Classification of solitary wave solutions}\label{sec:classify}

Rather than systematically work out all solutions $(\V(\z),\H(\z),\Y(\z))$ 
produced by the polynomial form \eqref{G.ansatz} for $G(\V)$, 
it will be more effective to proceed by imposing conditions so that 
only solutions of physical interest will be obtained. 
This will considerably simplify the necessary steps to find $(\V(\z),\H(\z),\Y(\z))$. 

As discussed in section~\ref{sec:intro},
the solutions that will be of most physical interest 
describe phase-modulated solitary waves on either a zero background or a non-zero background. 
In particular, both $\H(\z)$ and $\V(\z)$ should have an exponentially localized profile 
that asymptotically approaches a constant as $|\z|$ goes to infinity. 

The conditions under which such solutions will arise 
can be derived explicitly by studying the ODE \eqref{V''.G.reduc} for $\V(\z)$ 
appearing in Theorem~\ref{thm:reduction}. 
This ODE has the form of a nonlinear oscillator energy equation:
\begin{equation}\label{energy.eqn}
\tfrac{1}{2} \V'{}^2 + V(\V) =\E , 
\end{equation}
where, from the polynomial form \eqref{G.ansatz} for $G(\V)$, 
\begin{equation}\label{energy.potential}
V(\V) = s_2 \tfrac{1}{3}g_3 \V^3 +(s_2 \tfrac{1}{2}g_2 -\tfrac{1}{2}c) \V^2 +(s_2 g_1 -C_1) \V 
\end{equation}
represents the potential energy, 
and 
\begin{equation}\label{energy.E}
\E=\tilde C_4 -s_2 g_0
\end{equation}
represents the total energy. 
Consequently, 
existence of solitary wave solutions for $\V(\z)$ can be examined
by using a standard energy analysis method applied to the potential \eqref{energy.potential}. 

\begin{figure}[h!]
\centering
\includegraphics[trim=2cm 17cm 9cm 2cm,clip,width=0.48\textwidth]{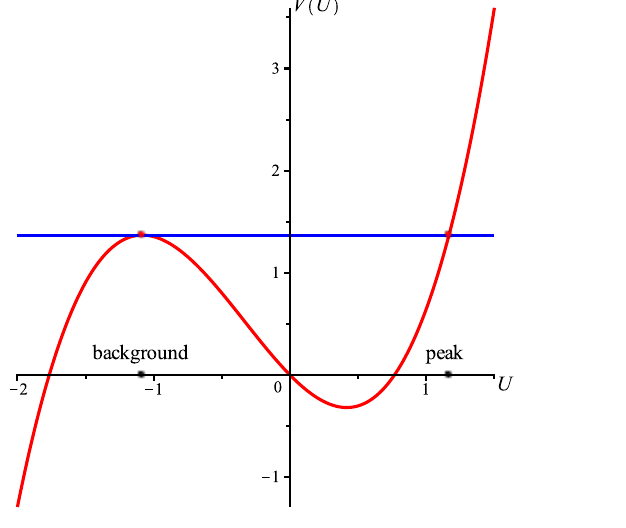}
\includegraphics[trim=2cm 17cm 9cm 2cm,clip,width=0.48\textwidth]{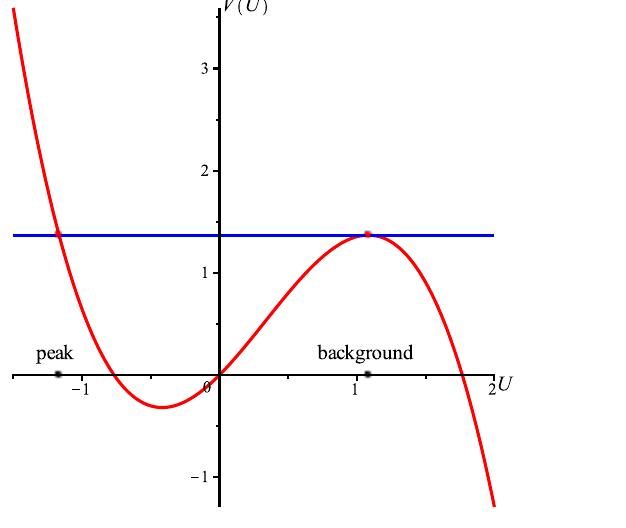}
\caption{Shape of potential for bright/dark solitary wave solutions}
\label{fig:up-down-potential}
\end{figure}

To begin, 
observe that the potential is a cubic polynomial, 
which is independent of the nonlinearity power $p$. 
The conditions for such a potential to support 
solitary waves on zero and non-zero backgrounds 
are well known from the cubic potential for travelling waves of the KdV equation:
$V(\V)$ must have both a local minimum and a local maximum. 
This requires that the discriminant of $V'(\V)$ is positive, 
\begin{equation}\label{solitary.cond}
d = (c- s_2 g_2)^2 +4 (s_2 C_1 -g_1) g_3 >0 . 
\end{equation}
The resulting shape of the potential is shown in Fig.~\ref{fig:up-down-potential}. 

\begin{prop}
The potential \eqref{energy.potential} supports a single solitary wave,
with $E$ being the local maximum of the potential. 
The corresponding solution for $\V(\z)$ is given by 
\begin{equation}\label{U.sol}
\V(\z) = b + s_3 h\, \sech^2\big(\sqrt{a}\, \z\big) , 
\end{equation}
where
\begin{equation}
h = \tfrac{3}{2}\sqrt{d}/|g_3| , 
%g=\tfrac{1}{6}|g_3| h,
\quad
a =\tfrac{1}{4}\sqrt{d}, 
%h=\tfrac{1}{2}\sqrt{d}/g,
\quad
b =-s_3\tfrac{1}{2}(\sqrt{d} -c +s_2 g_2),
\quad
E =(\tfrac{1}{2}\sqrt{d} +s_3 \tfrac{1}{3}|g_3|) b^2 , 
\end{equation}
and
\begin{equation}\label{s3}
s_3=s_2\,\sgn(g_3) =\pm1 . 
\end{equation}
This solitary wave \eqref{U.sol} will be 
\emph{bright} (namely, a positive peak) when $s_3=+1$, 
or \emph{dark} (namely, a negative peak) when $s_3=-1$. 
Its background is $b$ and its height/depth relative to the background is $h>0$. 
Its width is proportional to $w=2/\sqrt{a}$, with $a>0$. 
The height/depth and width are related by
$hw^2 = 12/|g_3|$.
\end{prop}

\begin{figure}[h!]
\centering
\includegraphics[trim=2cm 17cm 9cm 2cm,clip,width=0.6\textwidth]{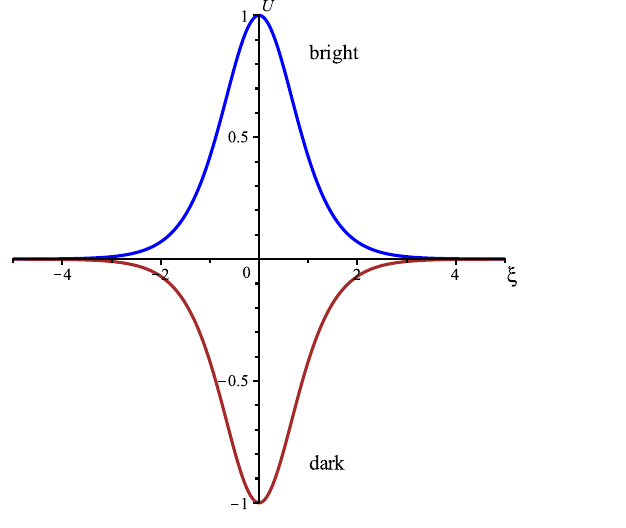}
\caption{Bright ($s_3=1$) and dark ($s_3=-1$) peaks for $\V(\z)$ on zero background}
\label{fig:U.brightdark} 
\end{figure}

Fig.~\ref{fig:U.brightdark} illustrates bright and dark peaks for $\V(\z)$. 

Next, from Theorem~\ref{thm:reduction}, 
the solutions for $\H(\z)$ and $\Y(\z)$ respectively are given by 
\begin{equation}\label{A.sol}
\H(\z) = \sqrt{ g_1 + g_2 \V(\z) + g_3 \V(\z)^2 - s_1 s_2\tfrac{1}{p+1} \V(\z)^{p+1} }
\end{equation}
and
\begin{equation}\label{Psi.sol}
\Y(\z)  = \frac{c}{2} \z \pm \sqrt{\mu} \int_0^\z \frac{d\z}{\H(\z)^2} + \phi , 
\end{equation}
where 
\begin{equation}
\phi 
\quad\text{ and }\quad
\mu = C_2{}^2
\end{equation}
are constants. 
When $\V(\z)$ has a localized profile, then so does $\H(\z)$,
and thus the resulting solution will describe a solitary wave. 

Some conditions are required for expressions \eqref{A.sol}--\eqref{Psi.sol} 
to be well-defined. 

First, the square-root \eqref{A.sol} must have 
\begin{equation}\label{Hsq.nonneg}
g_1 + g_2 \V(\z) + g_3 \V(\z)^2 - s_1 s_2\tfrac{1}{p+1} \V(\z)^{p+1} \geq 0
\end{equation}
for all $\z$. 
Since the extremal values of $\V(\z)$ are $\V(\pm\infty) = b$ and $\V(0) = b + s_3 h$, 
necessary conditions are given by 
$\H(\pm\infty)^2 \geq0$ and  $\H(0)^2 \geq0$,
where
\begin{equation}\label{Hsq}
\begin{aligned}
\H(\pm\infty)^2 = & 
g_1 + g_2 b + g_3 b^2 - s_1 s_2\tfrac{1}{p+1} b^{p+1} , 
\\
\H(0)^2 = & 
g_1 + g_2 (b+s_3 h) + g_3 (b+s_3 h)^2 - s_1 s_2\tfrac{1}{p+1} (b+s_3 h)^{p+1} . 
\end{aligned}
\end{equation}

Second, if $\mu\neq0$, 
convergence of the integral \eqref{Psi.sol} at $\z=0$ requires $\H(0)^2>0$,
since otherwise $\H(0)=0$ implies $\H(\z)^2 \sim O(\z^2)$ as $\z\sim0$,
which leads to a singularity at $\z=0$. 
Likewise, 
convergence for $|\z|\to\infty$ requires $\H(\pm\infty)^2>0$,
since otherwise a singularity occurs due to $\H(\z)^2 \sim O(e^{-a|\z|})$ as $\z\sim \pm\infty$. 

Therefore, the conditions 
\begin{subequations}\label{Hsq.conds}
\begin{align}
\H(\pm\infty)^2 >0, 
\quad
\H(0)^2 >0, 
\quad\text{ if } \mu\neq 0, 
\label{Hsq.conds.C2sqnot0}
\\
\H(\pm\infty)^2 \geq 0, 
\quad
\H(0)^2 \geq 0, 
\quad\text{ if } \mu= 0, 
\label{Hsq.conds.C2sqis0}
\end{align}
\end{subequations}
together with 
\begin{equation}\label{V.conds}
h>0,
\quad
a>0, 
\quad
\mu \geq 0
\end{equation}
are necessary. 
Note that if $\mu=0$ and $\H(0)=\H(\pm\infty)=0$ then the full condition \eqref{Hsq.nonneg} must be checked. 

Now, the final step will be to solve the second-order ODE \eqref{G.eqn} 
with $G(\V)$ having the polynomial form \eqref{G.ansatz}. 
Hereafter, it will be convenient to re-parameterize this polynomial by 
using the solitary wave parameters $b,h,a,E$, and $s_3=\pm1$:
\begin{equation}\label{g.rels}
g_0 = s_2 (\tilde C_4 - E),
\quad
g_1 = s_2 (C_1 +2 b h a +s_3 3 b^2 a) , 
\quad
g_2 = s_2 (c -2 (h  -s_3 3 b) a) , 
\quad
g_3 = s_2 s_3 3 a . 
\end{equation}
It will also be useful to put 
\begin{equation}\label{w.rel}
\w = \varpi - \tfrac{1}{4}c^2
\end{equation}
in terms of a modified frequency parameter $\varpi$. 
Substitution of the re-parameterized polynomial expression \eqref{G.ansatz} and \eqref{g.rels}--\eqref{w.rel}
into the ODE \eqref{G.eqn} yields a polynomial equation in $\V$ with powers
$1$, $2$, $3$, $4$, $5$, $p$, $p+1$, $p+2$, $p+3$, $p+4$, $2p$, $2p+1$, $2p+2$, $2p+3$, 
%$\V$, $\V^2$, $\V^3$, $\V^4$, $\V^5$, $\V^p$, $\V^{p+1}$, $\V^{p+2}$, $\V^{p+3}$, $\V^{p+4}$, $\V^{2p}$, $\V^{2p+1}$, $\V^{2p+2}$, $\V^{2p+3}$
and with coefficients that involve the constants 
$C_1,\tilde C_3,E$ (cf \eqref{energy.E}), 
the solitary wave parameters $b,h,a$, 
the physical constants $p,k,\varpi,c,\mu$,
and the signs $s_1,s_2,s_3$,
which are subject to the inequalities $p\neq0$, $k\neq0$, $c\neq0$, $h\neq0$, $a\neq0$
as well as the sign conditions \eqref{s1.s2} and \eqref{s3}. 

To solve the resulting polynomial equation, 
it must be split with respect to all distinct powers of $\V$. 
Since $p$ is an unknown, 
case distinctions arise when any two powers with at least one involving $p$ are equal. 
In each such case, 
the splitting gives an overdetermined algebraic system for the unknowns
$C_1,\tilde C_3,E,p,k,b,h,a,c,\varpi,\mu$, and $s_1,s_2,s_3$. 
All of the case distinctions and subsequent splitting and solving of the algebraic system 
have been carried out by using the computer algebra package Crack \cite{Wol}.
Further computational remarks are summarized in Appendix~\ref{sec:computation}. 

This computational analysis shows that solutions exist only for $p=1, 2, 3, 4$. 
For each case, 
explicit expressions for $g_1$, $g_2$, $g_3$, $E$,
and the determined parameters among $b,h,a,\mu,\varpi,c,k$ 
and also the signs $s_1,s_2,s_3$, 
are given in Appendix~\ref{sec:output}. 

Simplification of the solutions 
followed by analysis of the conditions \eqref{Hsq.conds} and \eqref{V.conds}
has been done in Maple. 
This step is highly non-trivial and relies heavily on the use of 
the parameterization in terms of $b,h,a,c,\varpi,\mu$. 
A key aspect is finding an optimal subset of these parameters for each solution 
such that, firstly, there are no unsolved algebraic equations,
and secondly, conditions \eqref{Hsq.conds} and \eqref{V.conds} 
reduce to a simple form characterized by 
a minimal set of inequalities holding on the free parameters. 
In some cases, a case splitting arises from this reduction, 
through the two subcases $\mu\neq0$ and $\mu=0$ in the conditions \eqref{Hsq.conds}, 
since the solution of the inequalities can have disconnected components when $\mu=0$.
The best outcome turns out to be achieved by 
giving a lower priority for $b$ to be solved for among the set of parameters 
rather than assuming that $c$ should be an arbitrary free parameter. 
This analysis is shown in Appendix~\ref{sec:output}. 

A classification of the resulting set of solution families is summarized in Table~\ref{table:soln.cases}. 
These solution families will be listed in the subsequent four subsections. 
Every family determines a corresponding explicit 
gKdV-LS solution \eqref{U.sol}--\eqref{Psi.sol} for $(\V(\z),\H(\z),\Y(\z))$, 
which are presented and discussed in section~\ref{sec:features}. 

A final remark here is that the lack of solutions for any $p>4$ is likely due to 
the limitation of considering only a polynomial form \eqref{G.ansatz} for $G(\V)$. 
It should be possible to extend the computational analysis to 
a rational form for $G(\V)$,
which may yield additional solutions, possibly involving higher powers for $p$. 
Such a form could also be relevant for seeking non-solitary wave solutions of the ODE system \eqref{U.A.Phi.sys},
such as heavy-tail solutions which have a power decay for large $|\z|$
\cite{AncNayRec}.

\begin{table}[H]
%\hbox{\hspace{-0.4in}
\begin{tabular}{|c|c|c||l|l|c||l|}
\hline
$p$
& $b$
& $\mu$
& free 
& determined
& no. of 
& label
\\
& 
&
&
&
& families 
&
\\
\hline
\hline
$4$ 
% Maple:pis4
& non-zero %$\neq 0$
& non-zero %$\neq0$
& $k$
& $b$, $h$, $a$, $\mu$, $\varpi$, $c$; $s_1$, $s_2$, $s_3$
& $1$
& 4-i
\\
\hline
$3$ 
% Maple:pis3
& non-zero % $\neq0$
& non-negative % $\geq0$
& $k$, $b$
& $h$, $a$, $\mu$, $\varpi$, $c$; $s_1$, $s_2$, $s_3$
& $1$
& 3-i
\\
% Maple:pis3-C2sqis0
& non-zero % $\neq 0$
& zero % $0$
& $k$
& $b$, $h$, $a$, $\mu$, $\varpi$, $c$; $s_1$, $s_2$, $s_3$
& $3$
& 3-i.1
\\
\hline
$2$ 
% Maple:pis2-case1
& arbitrary
& non-zero %$\neq0$
& $k$, $b$, $c$
& $h$, $a$, $\mu$, $\varpi$; $s_1$, $s_2$, $s_3$
& $1$
& 2-i
\\
% Maple:pis2-case1-C2sqis0-subcase2
& arbitrary 
& zero %$0$
& $k$, $b$
& $h$, $a$, $\mu$, $\varpi$, $c$; $s_1$, $s_2$, $s_3$
& $3$
& 2-i.1
\\
% Maple:pis2-case1-C2sqis0-subcase1
& non-zero %$\neq0$
& zero %$0$
& $k$, $c$
& $b$, $h$, $a$, $\mu$, $\varpi$; $s_1$, $s_2$, $s_3$
& $1$
& 2-i.2
\\
% Maple:pis2-case2
& non-zero %$\neq 0$
& non-zero % $\neq0$
& $b$, $\varpi$
& $k$, $h$, $a$, $\mu$, $c$; $s_1$, $s_2$, $s_3$
& $1$
& 2-ii
\\
% Maple:pis2-case2-C2sqis0
& non-zero %$\neq0$
& zero %$0$
& $b$
& $k$, $h$, $a$, $\mu$, $\varpi$, $c$; $s_1$, $s_2$, $s_3$
& $2$
& 2-ii.1
\\
\hline
$1$ 
% Maple:pis1-case1,pis1-case1-C2sqis0-subcase1
& arbitrary
& non-negative %$\neq0$
& $k$, $b$, $c$, $\varpi$
& $a$, $h$, $\mu$; $s_1$, $s_2$, $s_3$
& $1$
& 1-i
\\
% Maple:pis1-case1-C2sqis0-subcase2,pis1-case1-C2sqis0-subcase3
& arbitrary
& zero %$=0$
& $k$, $b$, $c$
& $a$, $h$, $\mu$, $\varpi$; $s_1$, $s_2$, $s_3$
& $2$
& 1-i.1
\\
% Maple:pis1-case3,pis1-case3-C2sqis0-subcase1
& arbitrary 
& non-negative %$\geq0$
& $b$, $a$, $c$, $\varpi$; $s_2$
& $k$, $h$, $\mu$; $s_1$, $s_3$
& $1$
& 1-ii
\\
% Maple:pis1-case3-C2sqis0-subcase2
& arbitrary 
& zero %$0$
& $b$, $c$, $\varpi$; $s_2$
& $k$, $a$, $h$, $\mu$; $s_1$, $s_3$
& $1$
& 1-ii.1
\\
% Maple:pis1-case4 
& arbitrary
& non-zero %$\neq0$ 
& $b$, $a$, $c$
& $k$, $h$, $\mu$, $\varpi$; $s_1$, $s_2$, $s_3$
& $1$
& 1-iii
\\
% Maple:pis1-case4-C2sqis0-subcase2,pis1-case4-C2sqis0-subcase3
& arbitrary %$\neq0$ 
& zero %$0$ 
& $b$, $c$
& $k$, $h$, $a$, $\mu$, $\varpi$; $s_1$, $s_2$, $s_3$
& $2$
& 1-iii.1
\\
% Maple:pis1-case4-C2sqis0-subcase1
& non-zero %$\neq0$ 
& zero %$0$ 
& $b$, $a$
& $k$, $h$, $\mu$, $\varpi$, $c$; $s_1$, $s_2$, $s_3$
& $1$
& 1-iii.2
\\
\hline
\hline
\end{tabular}
%}
\caption{Solution families}
\label{table:soln.cases}
\end{table}

\subsection{Solutions for $\boldsymbol{p=1}$}\label{sec:pis1.params}

\underline{Solution family (1-i)} has:\\
%maple pis1-case1 and pis1-case1-C2sqis0-subcase1
\noindent
signs
\begin{subequations}\label{pis1.fam1.soln}
\begin{equation}
s_1=1 ,
\quad
s_2 =\sgn(k-1/2),
\quad
s_3=\sgn(k) ; 
\end{equation}
background $b$, width and height of $\V$
\begin{gather}
w = \frac{2}{\sqrt{a}},
\quad
h = \frac{6a}{|k|},
\quad
a = \frac{kc+(2k-1)\varpi}{4(1-k)} - \frac{bk}{2} ;
\end{gather}
parameters for amplitude $\H$ and phase $\Y$
\begin{gather}
g_1 = \frac{|2k - 1|}{k^2}\Big( \frac{4k^2(c + \varpi)^2 +3(kc + \varpi)^2}{8(k - 1)^2} 
-\frac{6 a^2}{k^2} \Big), 
\\
g_2 = \frac{|2k - 1|(c + \varpi)}{k - 1},
\quad
g_3 = \sgn(k-1/2)\; k ,
\\
\mu =
\frac{4(2k - 1)^2}{k^4} (kb+\varpi) (kb +\varpi +4a)^2 (kb +\varpi +a)^2 \geq 0;
\end{gather}
speed $c$ and frequency 
\begin{equation}
\w=\varpi -\frac{c^2}{4} ;
\end{equation}
\end{subequations}
where the solution parameters $k\neq0$, $b$, $c$, $\varpi$
are required to satisfy the inequalities
\begin{subequations}\label{pis1.fam1.ineqns}
\begin{gather}
k\neq \frac{1}{2}, 1
\\
\varpi \geq -kb , 
\end{gather}
and
\begin{equation}\label{pis1.fam1.c}
\begin{aligned}
& c > 2(1 - k)b +\frac{(1-2k)\varpi}{k}
\text{ if } 0 < k < 1 ,
  \\
& c < 2(1 - k)b +\frac{(1-2k)\varpi}{k}
\text{ if } k<0 \text{ or } k > 1 . 
\end{aligned}
\end{equation}
\end{subequations}

The peak and tail expressions \eqref{Hsq} for $\H$ in this family are given by 
\begin{equation}\label{pis1.fam1.Hsq.muisnot0}
\begin{aligned}
\H(0) & = 
\sqrt{2|2k-1|}\frac{|kb+\varpi + a|}{|k|} 
> 0,
\\
\H(\pm\infty) & = 
\frac{\sqrt{2|2k-1|}}{|k|} \sqrt{ (kb+\varpi + a)(kb+\varpi + 4a) }
> 0. 
\end{aligned}
\end{equation}

The first of the two \underline{solution families (1-i.1)} has:\\
%maple pis1-case1-Csqis0-subcase2
\noindent
signs
\begin{subequations}\label{pis1.fam1.muis0.subcase2.soln}
\begin{equation}
s_1=1 ,
\quad
s_2 =\sgn(k-1/2),
\quad
s_3=\sgn(k) ; 
\end{equation}
background $b$, width and height of $\V$
\begin{gather}
w = \frac{2}{\sqrt{a}},
\quad
h = \frac{6a}{|k|},
\quad
a = \frac{c-b}{4} ;
\end{gather}
parameters for amplitude $\H$ and phase $\Y$
\begin{gather}
g_1 = \frac{|2k - 1|b^2}{2}, 
\quad
g_2 = -|2k - 1|b, 
\quad
g_3 = \sgn(k-1/2)\; k ,
\quad
\mu = 0; 
\end{gather}
speed $c$ and frequency 
\begin{equation}
\w=-(k-1)b - \frac{c(c+4)}{4} ;
\end{equation}
\end{subequations}
where the solution parameters $k\neq0$, $b$, $c$ 
are required to satisfy the inequalities
\begin{subequations}\label{pis1.fam1.muis0.subcase2.ineqns}
\begin{gather}
k\neq \frac{1}{2} , 
\\
c>b . 
\end{gather}
\end{subequations}

The peak and tail expressions \eqref{Hsq} for $\H$ are given by 
\begin{equation}\label{pis1.fam1.muis0.subcase2.Hsq}
\H(0) = 
\frac{3(c-b)}{2|k|}\sqrt{|2k-1|}
> 0,
\quad
\H(\pm\infty) = 0 . 
\end{equation}

The second of the \underline{solution families (1-i.1)} has:\\
%maple pis1-case1-Csqis0-subcase3
\noindent
signs
\begin{subequations}\label{pis1.fam1.muis0.subcase3.soln}
\begin{equation}
s_1=1 ,
\quad
s_2 =-\sgn(k-1/2),
\quad
s_3=\sgn(k) ; 
\end{equation}
background $b$, width and height of $\V$
\begin{gather}
w = \frac{2}{\sqrt{a}},
\quad
h = \frac{6a}{|k|},
\quad
a = \frac{k(c-b)}{2k-3} ;
\end{gather}
parameters for amplitude $\H$ and phase $\Y$
\begin{gather}
g_1 = |2k - 1|b \Big( \frac{3c}{2k-3} -\frac{b}{2} \Big), 
\quad
g_2 = |2k - 1|\frac{2kb -3c}{2k-3}, 
\quad
g_3 = -\sgn(k-1/2)\; k ,
\quad
\mu = 0; 
\end{gather}
speed $c$ and frequency 
\begin{equation}
\w=\frac{k(c - 2(k-1)b)}{2k-3} - \frac{c^2}{4} ;
\end{equation}
\end{subequations}
where the solution parameters $k\neq0$, $b$, $c$ 
are required to satisfy the inequalities
\begin{subequations}\label{pis1.fam1.muis0.subcase3.ineqns}
\begin{gather}
k\neq \frac{1}{2} , 
\\
\begin{aligned}
& c > b
\text{ if } 0 < k < 3/2 ,
  \\
& c < b
\text{ if } k<0 \text{ or } k > 3/2 . 
\end{aligned}
\end{gather}
\end{subequations}

The peak and tail expressions \eqref{Hsq} for $\H$ vanish, 
\begin{equation}\label{pis1.fam1.muis0.subcase3.Hsq}
\H(0) = \H(\pm\infty) = 0 .
\end{equation}

\underline{Solution family (1-ii)} has:\\
%maple pis1-case3
\noindent
interaction coefficient
\begin{subequations}\label{pis1.fam2.soln}
\begin{equation}
k = \frac{1}{6} ;
\end{equation}
signs
\begin{equation}
s_1=1,
\quad
s_2=\pm 1,
\quad
s_3=1 ; 
\end{equation}
background $b$, width and height of $\V$
\begin{gather}
w = \frac{2}{\sqrt{a}},
\quad
h = 12 a ;
\end{gather}
parameters for amplitude $\H$ and phase $\Y$
\begin{gather}
g_1 = s_2 3(b - c + 4a)(b + 4a + 4\varpi), 
\quad
g_2 = -s_2 (b - c + 4a), 
\quad
g_3 = s_2\frac{1}{2},  
\\
\mu = \frac{2(b - c + 4a)^2 (b + 6(\varpi+a))^2 (b + 6\varpi)}{3} \geq 0;
\end{gather}
speed $c$ and frequency 
\begin{equation}
\w=\varpi -\frac{c^2}{4} ;
\end{equation}
\end{subequations}
where the solution parameters $a>0$, $b$, $c$, $\varpi$
are required to satisfy the inequalities
\begin{subequations}\label{pis1.fam2.ineqns}
\begin{equation}\label{pis1.fam2.ineqn1}
\varpi \geq -\frac{b}{6} , 
\end{equation}
and
\begin{equation}\label{pis1.fam2.ineqn2}
\begin{aligned}
& c < b + 4a
\text{ if } s_2 =1 , 
\\
& c > b + 4a
\text{ if } s_2 =-1 .  
\end{aligned}
\end{equation}
\end{subequations}

The peak and tail expressions \eqref{Hsq} for $\H$ are given by 
\begin{equation}\label{pis1.fam2.Hsq}
\H(0) = \sqrt{2(b+6\varpi)|b-c+4a|},
\quad
\H(\pm\infty) = \sqrt{2(b+6(\varpi+a))|b-c+4a|} . 
\end{equation}

\underline{Solution family (1-ii.1)} has:\\
%maple pis1-case3-C2sqis0-subcase2
\noindent
interaction coefficient
\begin{subequations}\label{pis1.fam2.muis0.subcase2.soln}
\begin{equation}
k = \frac{1}{6} ;
\end{equation}
signs
\begin{equation}
s_1=1,
\quad
s_2=\pm 1,
\quad
s_3=1 ; 
\end{equation}
background $b$, width and height of $\V$
\begin{gather}
w = \frac{2}{\sqrt{a}},
\quad
h = 12 a,
\quad
a = -\varpi - \frac{b}{6}; 
\end{gather}
parameters for amplitude $\H$ and phase $\Y$
\begin{gather}
g_1 = s_2 \frac{b(b - 3c -12\varpi)}{3}, 
\quad
g_2 = -s_2 \frac{b - 3c -12\varpi}{3}, 
\quad
g_3 = s_2\frac{1}{2},  
\quad
\mu = 0;
\end{gather}
speed $c$ and frequency 
\begin{equation}
\w=\varpi -\frac{c^2}{4} ;
\end{equation}
\end{subequations}
where the solution parameters $a>0$, $b$, $c$, $\varpi$
are required to satisfy the inequalities
\begin{subequations}\label{pis1.fam2.muis0.subcase2.ineqns}
\begin{equation}\label{pis1.fam2.muis0.subcase2.ineqn1}
\varpi > -\frac{b}{6}, 
\end{equation}
and
\begin{equation}\label{pis1.fam2.muis0.subcase2.ineqn2}
\begin{aligned}
& c > \frac{b}{3} -4\varpi 
\text{ if } s_2 =1 , 
\\
& c < \frac{b}{3} -4\varpi 
\text{ if } s_2 =-1 .  
\end{aligned}
\end{equation}
\end{subequations}

In this family, the peak and tail expressions \eqref{Hsq} for $\H$ are given by 
\begin{equation}\label{pis1.fam2.muis0.subcase2.Hsq}
\H(0) = \sqrt{2(b+6\varpi)\Big|\frac{b}{3}- c -4 \varpi\Big|},
\quad
\H(\pm\infty) = 0 . 
\end{equation}

\underline{Solution family (1-iii)} has:\\
%maple pis1-case4
\noindent
interaction coefficient
\begin{subequations}\label{pis1.fam3.soln}
\begin{equation}
k = 1 ;
\label{pis1.fam3.k}
\end{equation}
signs
\begin{equation}
s_1=1,
\quad
s_2 =1,
\quad
s_3=1 ; 
\end{equation}
background $b$, width and height of $\V$
\begin{gather}
w = \frac{2}{\sqrt{a}},
\quad
h = 6 a ;
\end{gather}
parameters for amplitude $\H$ and phase $\Y$
\begin{gather}
g_1 = \frac{7(7b - 8c + 8a)(7b - 2c + 8a)}{4} -\frac{3b(7b -10c)}{8}, 
\quad
g_2 = c-2b -4a,
\quad
g_3 = 1, 
\\
\mu =4(b -c)(b -c +a)^2(b -c + 4a)^2 >0; 
\end{gather}
speed $c$ and frequency 
\begin{equation}
\w=-\frac{c(c+4)}{4} ;
\end{equation}
\end{subequations}
where the solution parameters $a>0$, $b$, $c$ are required to satisfy the inequality
\begin{equation}\label{pis1.fam3.ineqns}
c < b . 
\end{equation}

The peak and tail expressions \eqref{Hsq} for $\H$ are given by 
\begin{equation}\label{pis1.fam3.Hsq}
\H(0) = \sqrt{2(b -c + a)}, 
\quad
\H(\pm\infty) = \sqrt{2(b -c +a)(b -c +4a)} . 
\end{equation}

The first \underline{solution family (1-iii.1)} has:\\
%maple pis1-case4-C2sqis0-subcase2
\noindent
interaction coefficient
\begin{subequations}\label{pis1.fam3.C2sqis0.subcase2.soln}
\begin{equation}
k = 1 ;
\end{equation}
signs
\begin{equation}
s_1=1,
\quad
s_2 =1,
\quad
s_3=1 ; 
\end{equation}
background $b$, width and height of $\V$
\begin{gather}
w = \frac{2}{\sqrt{a}},
\quad
h = 6 a ,
\quad
a = \frac{c-b}{4} ;
\end{gather}
parameters for amplitude $\H$ and phase $\Y$
\begin{gather}
g_1 = \frac{b^2}{2}, 
\quad
g_2 = -b, 
\quad
g_3 = 1, 
\quad
\mu = 0; 
\end{gather}
speed $c$ and frequency 
\begin{equation}
\w=-\frac{c(c+4)}{4} ;
\end{equation}
\end{subequations}
where the solution parameters $b$, $c$ are required to satisfy the inequality
\begin{equation}\label{pis1.fam3.C2sqis0.subcase2.ineqns}
c > b . 
\end{equation}

The peak and tail expressions \eqref{Hsq} for $\H$ are given by 
\begin{equation}\label{pis1.fam3.muis0.subcase2.Hsq}
\H(0) = \frac{3}{2}\sqrt{\frac{c-b}{2}}, 
\quad
\H(\pm\infty) = 0 . 
\end{equation}

The second \underline{solution family (1-iii.1)} has:\\
%maple pis1-case4-C2sqis0-subcase3
\noindent
interaction coefficient
\begin{subequations}\label{pis1.fam3.C2sqis0.subcase3.soln}
\begin{equation}
k = 1 ;
\end{equation}
signs
\begin{equation}
s_1=1,
\quad
s_2 =-1,
\quad
s_3=1 ; 
\end{equation}
background $b$, width and height of $\V$
\begin{gather}
w = \frac{2}{\sqrt{a}},
\quad
h = 6 a ,
\quad
a = c-b ;
\end{gather}
parameters for amplitude $\H$ and phase $\Y$
\begin{gather}
g_1 = \frac{b(5b -6c)}{2}, 
\quad
g_2 = 3c -2b, 
\quad
g_3 = -1, 
\quad
\mu = 0; 
\end{gather}
speed $c$ and frequency 
\begin{equation}
\w=-\frac{c(c+4)}{4} ;
\end{equation}
\end{subequations}
where the solution parameters $b$, $c$ are required to satisfy the inequality
\begin{equation}\label{pis1.fam3.C2sqis0.subcase3.ineqns}
c > b . 
\end{equation}

The peak and tail expressions \eqref{Hsq} for $\H$ vanish, 
\begin{equation}\label{pis1.fam3.muis0.subcase3.Hsq}
\H(0) = \H(\pm\infty) = 0 . 
\end{equation}

\underline{Solution family (1-iii.2)} has:\\
%maple pis1-case4-C2sqis0-subcase1
\noindent
interaction coefficient
\begin{subequations}\label{pis1.fam3.C2sqis0.subcase1.soln}
\begin{equation}
k = 1 ;
\end{equation}
signs
\begin{equation}
s_1=1,
\quad
s_2 =1,
\quad
s_3=1 ; 
\end{equation}
background $b$, width and height of $\V$
\begin{gather}
w = \frac{2}{\sqrt{a}},
\quad
h = 6 a ;
\end{gather}
parameters for amplitude $\H$ and phase $\Y$
\begin{gather}
g_1 = \frac{(b + 4a)^2}{2}, 
\quad
g_2 = -b -4a, 
\quad
g_3 = 1, 
\quad
\mu = 0; 
\end{gather}
speed and frequency 
\begin{equation}
c=b,
\quad
\w=-\frac{b(b+4)}{4} ;
\end{equation}
\end{subequations}
where the solution parameters are $b$, $a>0$.  

The peak and tail expressions \eqref{Hsq} for $\H$ are given by 
\begin{equation}\label{pis1.fam3.muis0.subcase1.Hsq}
\H(0) = \sqrt{2}a, 
\quad
\H(\pm\infty) = 2\sqrt{2}a .
\end{equation}

\subsection{Solutions for $\boldsymbol{p=2}$}\label{sec:pis2.params}

\underline{Solution family (2-i)} has:\\
%maple pis2-case1
\noindent
signs
\begin{subequations}\label{pis2.fam1.soln}
\begin{equation}
s_1= \sgn(k) ,
\quad
s_2=1,
\quad
s_3 = \sgn(k) ;
\end{equation}
background $b$, width and height of $\V$
\begin{equation}
w =\frac{2}{\sqrt{a}} ,
\quad
h = \frac{12a}{|k|},
\quad
a = \frac{|k| c}{8(|k| -b)} + \frac{(2b(b -2|k|) + k^2)k}{16(|k| -b)} ;
\end{equation}
parameters for amplitude $\H$ and phase $\Y$
\begin{gather}
g_1 = s_1\frac{|k| -2 b}{(|k| - b)^2}\Big( s_1 c +\frac{3|k|b}{16} -\frac{5k^2}{16} - \frac{3b^2}{8} \Big)^2
 -s_1\Big(|k| -\frac{22b}{3}\Big)\Big(\frac{3|k|}{16} - \frac{b}{8}\Big)^2,
\\
g_2 =  \frac{(|k| - 2b)(s_12c -k^2)}{4(|k| - b)} -\frac{kb}{2}, 
\quad
g_3 = \frac{k}{2} , 
\\
\mu = 8|k|(|k| - 2 b)^2 \Big(\frac{2a}{|k|} +\frac{5(k b -2 c)}{4(|k| - b)}\Big)
\Big(\frac{2a}{|k|} +\frac{5(k b -2 c)}{8(|k| - b)}\Big)^2
\Big(\frac{2a}{|k|} +\frac{k b -2 c}{4(|k| - b)}\Big)^2 >0;
\end{gather}
speed $c$ and frequency 
\begin{equation}
\w=-\frac{c^2}{4} -bk + \frac{k(2 b^2 +6|k|b +k^2) -18|k| c)}{16(|k| - b)} ;
\end{equation}
\end{subequations}
where the solution parameters $k\neq0$, $b$, $c$ are required to satisfy the inequalities 
\begin{subequations}\label{pis2.fam1.ineqns}
\begin{equation}\label{pis2.fam1.kpos.ineqn}
b < k/2,
\quad
-b^2 - k^2/2 + 2k b < c \leq b^2/9 + k^2/18 + k b/3 
,
\quad
\text{ if } k>0 , 
\end{equation}
and
\begin{equation}\label{pis2.fam1.kneg.ineqn}
\begin{aligned}
|k|/2 < b < |k|,
&\quad
b^2 + k^2/2 - 2 |k| b < c \leq -b^2/9 - k^2/18 -|k| b/3 
\\
b > |k|,
&\quad
-b^2/9 -k^2/18 -|k| b/3 \leq c < b^2 +k^2/2 -2 |k| b 
\end{aligned}
\quad
\text{ if } k<0
\end{equation}
\end{subequations}

The peak and tail expressions \eqref{Hsq} for $\H$ in this family are given by 
\begin{equation}\label{pis2.fam1.Hsq.muisnot0}
\begin{aligned}
\H(0) & = 
\sqrt{ \frac{5k(|k|-2b) - 72a}{|k|} }\; \frac{\big|k(|k|-2b) -8a\big|}{|k|}
> 0,
\\
\H(\pm\infty) & = 
\sqrt{ \frac{|k|-2b}{8k|k|} \big(5k(|k|-2b) - 64a\big) \big(k(|k|-2b) - 8a\big) }
> 0 . 
\end{aligned}
\end{equation}

The three \underline{solution families (2-i.1)} have:\\
%maple pis2-case1-C2sqis0-subcase2 
\noindent
values 
\begin{equation}
(\nu_1,\nu_2) = 
(1/8,1/8),
\quad
(1/3,1/18),
\quad
(-1,1/2) ;
\end{equation}
signs
\begin{subequations}\label{pis2.fam1.muis0.subcase2.soln}
\begin{equation}
s_1= 
\begin{cases}
\sgn(k) & \text{ for } (\nu_1,\nu_2) = (1/3,1/18), (-1,1/2) 
\\
-\sgn(k) & \text{ for } (\nu_1,\nu_2) = (1/8,1/8)
\end{cases}, 
\quad
s_2=1,
\quad
s_3 = \sgn(k) ;
\end{equation}
background $b$, width and height of $\V$
\begin{equation}
w =\frac{2}{\sqrt{a}} ,
\quad
h = \frac{12a}{|k|},
\quad
a = \frac{k}{8(k -s_1 b)}\Big( b k(\nu_1-2) + s_1(2\nu_2 +1)\Big(b^2 +\frac{k^2}{2}\Big) \Big);
\end{equation}
parameters for amplitude $\H$ and phase $\Y$
\begin{gather}
\begin{aligned}
g_1 = & -\frac{2s_1b- k}{(s_1 b -k)^2} \Big( \Big(2\nu_2 -\frac{3}{8}\Big) b^2 +\Big(2\nu_2 -\frac{5}{8}\Big) \frac{k^2}{2} +s_1 \Big(\nu_1+\frac{3}{16}\Big) bk \Big)^2 
\\&\qquad
+ \frac{(2s_1b -3 k)^2}{3(16(s_1 b -k))^2} \big( 22 s_1 b - 3k \big), 
\end{aligned}
\\
g_2 =  \frac{2s_1b- k}{2(s_1 b -k)} \big( s_1\nu_2 (2b^2 +k^2) + \nu_1 b k  \big) 
-\frac{s_1 k(2b^2-k^2)}{4(s_1b - k)} , 
% g_2 =  \frac{s_1(2\nu_2-1)}{4} ( (s_1 b -k)^2 +b^2 ) +\frac{b(s_1b-k)}{4} (2(\nu_2 +\nu_1) -3) -\frac{b^3}{4(s_1b -k)} ( 6\nu_2 + 2\nu_1 -1 )
\quad
g_3 = \frac{k}{2} , 
\\
\mu = 0 ; 
\end{gather}
speed 
\begin{equation}
c = \nu_1 b k + s_1 \nu_2 (2b^2+k^2) ; 
\end{equation}
and frequency 
\begin{equation}
\w=\frac{k}{8(s_1 b - k)} \Big( (9\nu_1 +5) bk +s_1\Big( 9(2\nu_2 -1)b^2 + (18\nu_1 -1)\frac{k^2}{2} \Big) \Big);
\end{equation}
\end{subequations}
where the solution parameters $k\neq0$, $b$ are required to satisfy the inequalities 
\begin{subequations}\label{pis2.fam1.muis0.subcase2.ineqns}
\begin{align}
& b < s_1 k/2
\quad
\text{ if } k>0 , 
\\
& b > -s_1|k|/2
\quad
\text{ if } k<0 . 
\end{align}
\end{subequations}

The peak and tail expressions \eqref{Hsq} for $\H$ in this family are given by 
\begin{equation}\label{pis2.fam1.muis0.subcase2.Hsq}
\begin{aligned}
\H(0) & = \begin{cases}
\dfrac{3}{64}\sqrt{5(s_1 2b -k)^3} , 
& (\nu_1,\nu_2)=(1/8,1/8) 
\\
0 , 
& (\nu_1,\nu_2)=(1/3,1/18), (-1,1/2) 
\end{cases} , 
%\sqrt{ \frac{\big( (18 \nu_2 -1)(2 b^2 + k^2) + s_1 6 (3\nu_1  - 1) b k  \big)}{(4 (s_1 b - k))^3} } \big| (2 \nu_2 -1)(2 b^2 + k^2) + s_1 2 (\nu_1  + 1) b k  \big|
\\
\H(\pm\infty) & = \begin{cases}
0 , & (\nu_1,\nu_2)=(1/8,1/8), (-1,1/2) 
\\
\dfrac{1}{18}\sqrt{10(k - s_1 2b)^3} , 
& (\nu_1,\nu_2)=(1/3,1/18) 
\end{cases} . 
%\sqrt{ \frac{(k - 2 s_1 b)}{(4 (s_1 b- k))^2} \big( (8 \nu_2 -1)(2 b^2 + k^2) + s_1 (8\nu_1  - 1) b k  \big) \big( (2 \nu_2 -1)(2 b^2 + k^2) + s_1 2 (\nu_1  + 1) b k  \big) }
\end{aligned}
\end{equation}

\underline{Solution family (2-i.2)} has:\\
%maple pis2-case1-C2sqis0-subcase1
\noindent
signs
\begin{subequations}\label{pis2.fam1.muis0.subcase1.soln}
\begin{equation}
s_1 = -\sgn(k) ,
\quad
s_2=1,
\quad
s_3 = \sgn(k) ;
\end{equation}
background, height, width of $\V$
\begin{equation}
b =-|k|/2 <0, 
\quad
h  = \frac{12 a}{|k|}, 
\quad
w = \frac{2}{\sqrt{a}},
\quad
a =\frac{k|k| +4c}{16} ; 
\end{equation}
parameters for amplitude $\H$ and phase $\Y$
\begin{gather}
g_1 = \frac{k^3}{24}, 
\quad
g_2 = \frac{k|k|}{4}, 
\quad
g_3 = \frac{k}{2} ,
\quad
\mu = 0 ; 
\end{gather}
speed $c$ and frequency 
\begin{equation}
\w = -\frac{c(c+9)}{4} - \frac{k|k|}{16} ;
\end{equation}
\end{subequations}
where the solution parameters $k\neq0$ and $c$ are required to satisfy the inequalities 
\begin{equation}\label{pis2.fam1.subcase1.ineqns}
c> -k|k| . 
\end{equation}

The peak and tail expressions \eqref{Hsq} for $\H$ are given by 
\begin{equation}\label{pis2.fam1.muis0.subcase1}
\H(0) = 3\sqrt{ \Big(\frac{k}{4} + \frac{c}{|k|}\Big)^3 },
\quad
\H(\pm\infty) = 0 . 
\end{equation}

\underline{Solution family (2-ii)} has:\\
%maple pis2-case2
\noindent
signs 
\begin{subequations}\label{pis2.fam2.soln}
\begin{equation}
s_1 =-1,
\quad
s_2=1,
\quad
s_3=-1 ;
\end{equation}
background, height, width of $\V$
\begin{equation}
b =|k| >0,
\quad
h  = \frac{12 a}{|k|}, 
\quad
w = \frac{2}{\sqrt{a}},
\quad
a =\frac{13k^2 - 4\varpi}{72} ;
\end{equation}
parameters for amplitude $\H$ and phase $\Y$
\begin{gather}
g_1 = \frac{61 |k|^3}{81} + \frac{64 \varpi^2}{81|k|} -\frac{127 |k|\varpi}{81}, 
\quad
g_2 = \frac{4\varpi}{9} - \frac{2 k^2}{9}, 
\quad
g_3 = -\frac{|k|}{2} ,
\\
\mu = \frac{\varpi - k^2}{2|k|}\Big( \frac{(2\varpi - k^2)(64\varpi - 59 k^2)}{81}\Big)^2 >0;
\end{gather}
frequency 
\begin{equation}
\w = \varpi - \frac{k^4}{16}
\end{equation}
and speed
\begin{equation}
c = -\frac{k^2}{2}<0 ;
\end{equation}
\end{subequations}
where the solution parameters $k\neq0$ and $\varpi$ are required to satisfy the inequalities 
\begin{subequations}\label{pis2.fam2.ineqns}
\begin{equation}
k<0 , 
\end{equation}
and 
\begin{equation}\label{pis2.fam2.varpi}
k^2 < \varpi < 13k^2/8 . 
\end{equation}
\end{subequations}

In this family, the peak and tail expressions \eqref{Hsq} for $\H$ are given by 
\begin{equation}\label{pis2.fam2.Hsq}
\H(0) = \frac{4}{9}(2\varpi -k^2)\sqrt{ \frac{\varpi - k^2}{|k|^3} }
>0,
\quad
\H(\pm\infty) = \frac{1}{18} \sqrt{ \frac{2(64\varpi - 59k^2)(2\varpi -k^2)}{|k|} }
>0 . 
\end{equation}

\underline{Solution family (2-ii.1)} has:\\
%maple pis2-case2-C2sqis0
signs 
\begin{subequations}\label{pis2.fam2.subcase1.soln}
\begin{equation}
s_1 =-1,
\quad
s_2=1,
\quad
s_3=\begin{cases}
1, & \text{ for } \lambda = 59/64 
\\
-1, & \text{ for } \lambda = 1/2
\end{cases} ;
\end{equation}
background, height, width of $\V$
\begin{equation}
b =-k ,
\quad
h  = \frac{12 a}{|k|}, 
\quad
w = \frac{2}{\sqrt{a}},
\quad
a =\frac{k^2(13 -8\lambda)}{72} ;
\end{equation}
parameters for amplitude $\H$ and phase $\Y$
\begin{gather}
g_1 = \frac{k^3(-64\lambda^2 +127\lambda -61)}{81}, 
\quad
g_2 = \frac{2k^2(2\lambda -1)}{9} ,
\quad
g_3 = \frac{k}{2} ,
\quad
\mu = 0 ;
\end{gather}
frequency 
\begin{equation}
\w = \frac{k^2(16\lambda -k^2)}{16}
\end{equation}
and speed
\begin{equation}
c = -\frac{k^2}{2}<0 ;
\end{equation}
\end{subequations}
where the solution parameter is $k\neq 0$. 

The tail and peak expressions \eqref{Hsq} for $\H$ are given by 
\begin{equation}\label{pis2.fam2.subcase1.Hsq}
\H(\pm\infty) = 0,
\quad
\H(0) = \begin{cases}
0, 
& \text{ for } \lambda =1/2 
\\
\dfrac{3\sqrt{10}}{32}, 
& \text{ for } \lambda =59/64
\end{cases} . 
\end{equation}

\subsection{Solutions for $\boldsymbol{p=3}$}\label{sec:pis3.params}

\underline{Solution family (3-i)} has:\\
%maple pis3
\noindent
signs
\begin{subequations}\label{pis3.soln}
\begin{equation}
s_1=1,
\quad
s_2 =-1,
\quad
s_3=\sgn(k) ;
\end{equation}
background $b$, width and height of $\V$
\begin{align}
w =\frac{2}{\sqrt{a}} ,
\quad
h = \frac{20a}{|k|},
\quad
a = \frac{|k|(k+3b^2)}{40|b|} ; 
\end{align}
parameters for amplitude $\H$ and phase $\Y$
\begin{gather}
g_1 =\frac{b^4}{4} -\frac{11k b^2}{15} + \frac{91k^2}{150} -\frac{k^3}{10 b^2} ,
\quad
g_2 =  \frac{(2k-b^2)(3k-5b^2)}{45 b}, 
\quad
g_3 = -\frac{3k}{10} , 
\\
\mu = \frac{4k(2k -b^2)(k-5b^2)^2(3k-5b^2)^2(15k- 11b^2)^2}{3(15b)^5} \geq 0;
\end{gather}
frequency 
\begin{equation}
\w=\frac{2k (k + 3b^2)}{5b}  - \Big(\frac{b^2}{18} - \frac{53 k}{180} + \frac{7 k^2}{60b^2} \Big)^2 ;
\end{equation}
and speed
\begin{equation}
c= \frac{7k(3b^2 -k)}{30b} -\frac{b(b^2 +k)}{9} ;
\end{equation}
\end{subequations}
where the solution parameters are $k\neq0$, $b$. 
These parameters are required to satisfy the inequalities
\begin{equation}\label{pis3.ineqns}
\begin{aligned}
& b<0, 
\quad
b^2 \geq 2k
&&
\text{ if } 
k>0 , 
\\
& b>0, 
\quad
b^2 > |k|/3
&&
\text{ if } 
k<0 . 
\end{aligned}
\end{equation}

The peak and tail expressions \eqref{Hsq} for $\H$ in this family are given by 
\begin{equation}\label{pis3.Hsq.muisnot0}
\H(0) = \frac{(5b^2 -k) (11b^2 -15k)}{120 b^2} > 0, 
\quad
\H(\pm\infty) = \frac{\sqrt{(5b^2 -k)(5b^2 -3k)(11b^2 -15k)}}{15\sqrt{2}|b|} > 0 . 
\end{equation}

The three \underline{solution families (3-i.1)} have:\\
%maple pis3-C2sqis0, cases 2,3,4
\noindent
values
\begin{subequations}\label{pis3.soln.muis0}
\begin{equation}
\lambda = 
\frac{1}{5},
\quad
\frac{3}{5},
\quad
\frac{15}{11} ; 
\end{equation}
signs
\begin{equation}
s_1=1,
\quad
s_2 =\begin{cases} 
1 & \text{ for } \lambda = 1/5, 15/11
\\
-1 & \text{ for } \lambda = 3/5
\end{cases},
\quad
s_3=1 ;
\end{equation}
background, width and height of $\V$
\begin{gather}
b = -\lambda\sqrt{k} <0, 
\quad
w = \frac{2\sqrt{\lambda}}{\sqrt{1+3\lambda^2}} \frac{1}{\sqrt[4]{k^3}} ,
\quad
h = \frac{(1+3\lambda^2)}{2\lambda} \sqrt{k} ; 
\end{gather}
parameters for amplitude $\H$ and phase $\Y$
\begin{gather}
g_1 =\Big( \frac{1}{10 \lambda^2} -\frac{91}{150} +\frac{11\lambda^2}{15} -\frac{\lambda^4}{4} \Big) k^2,
\quad
g_2 =  \frac{(\lambda^2 -2)(3-5\lambda^2)}{45 \lambda} \sqrt{k^3}, 
\quad
g_3 = \frac{3k}{10} , 
\quad
\mu = 0 ;
\end{gather}
speed
\begin{equation}
c= \Big(\frac{7(3\lambda^2 -1)}{30\lambda} -\frac{\lambda(1+\lambda^2)}{9}\Big)\sqrt{k^3} 
\end{equation}
and frequency 
\begin{equation}
\w=\frac{2(1- 3\lambda^2)}{5\lambda}\sqrt{k^3} - \frac{c^2}{4} ; 
\end{equation}
\end{subequations}
%\begin{equation}
%\lambda = \frac{1}{5}, \frac{15}{11}
%\end{equation}
where the solution parameters are $k\neq0$, $\lambda$, 
with $k$ required to satisfy the inequality
\begin{equation}
k>0 . 
\end{equation}

In these families, the tail and peak expressions \eqref{Hsq} are given by 
\begin{equation}
\begin{gathered}
\H(\pm\infty) = \H(0) = 0
\quad\text{ for $\lambda = \frac{1}{5}, \frac{15}{11}$ }, \\
\H(\pm\infty) =0,
\quad
\H(0)  = \dfrac{7k^2}{30}, 
\quad\text{ for $\lambda = \frac{3}{5}$ }.
\end{gathered}
\end{equation}

\subsection{Solutions for $\boldsymbol{p=4}$}\label{sec:pis4.params}

\underline{Solution family (4-i)} has:\\
%maple pis4
\noindent
signs 
\begin{subequations}\label{pis4.soln}
\begin{equation}
s_1 =1,
\quad
s_2=1,
\quad
s_3=1 ;
\end{equation}
background, height, width of $\V$
\begin{align}
b = -\sqrt[3]{2k} <0,
\quad
h = \frac{3\sqrt[3]{2k}}{2} ,
\quad
w = \frac{4\sqrt{10}}{\sqrt[3]{2k}^2} ;
\end{align}
parameters for amplitude $\H$ and phase $\Y$
\begin{gather}
g_1 = \frac{9\sqrt[3]{2k}^5}{25}, 
\quad
g_2 = -\frac{3\sqrt[3]{2k}^4}{20}, 
\quad
g_3 = \frac{k}{5}, 
\quad
\mu = 2\Big(\frac{81 \sqrt[3]{2k}^7}{400}\Big)^2 >0;
\end{gather}
frequency 
\begin{equation}
\w = \frac{\sqrt[3]{2k}^4 (40 -\sqrt[3]{2k}^4)}{64} ;
\end{equation}
and speed
\begin{equation}
c = -\frac{\sqrt[3]{2k}^4}{4} <0;
\end{equation}
\end{subequations}
where the solution parameter $k$ is required to satisfy the inequality
\begin{equation}\label{pis4.ineqns}
k>0 . 
\end{equation}

In this family, the tail and peak expressions \eqref{Hsq} for $\H$ are given by 
\begin{equation}\label{pis4.Hsq}
\H(\pm\infty) = \frac{9\sqrt[6]{2k}^5}{10} >0,
\quad
\H(0) = \sqrt{\frac{3}{8}}\H(\pm\infty) . 
\end{equation}

\section{Phase-modulated solitary waves}\label{sec:features}

From the results in section~\ref{sec:classifysolns}, 
the explicit solitary wave solutions \eqref{soln} of the gKdV-LS system \eqref{u.eqn}--\eqref{psi.eqn}
will now be presented and their kinematic features will be discussed. 
These solutions hold for the nonlinearity powers $p=1,2,3,4$,
covering all physically interesting cases for the nonlinearity in the gKdV-LS system.

The gKdV variable $u=\V(\z)$ is given by the sech-squared expression \eqref{U.sol}. 
Physically, it describes a solitary wave on a background $b$, 
with a height/depth $h>0$, 
where the wave is \emph{bright} when $s_3>0$ and \emph{dark} when $s_3<0$. 
Its width is proportional to $w=2/\sqrt{a}$ in terms of the parameter $a>0$. 

The LS variable $\psi=\H(\z)\exp(i(\w t+\Y(\z))$ is given by 
the amplitude expression \eqref{A.sol} and the phase expression \eqref{Psi.sol}. 
These expressions exhibit a wide variety of features. 

\begin{figure}[h!]
\centering
\includegraphics[trim=2cm 17cm 9cm 2cm,clip,width=0.48\textwidth]{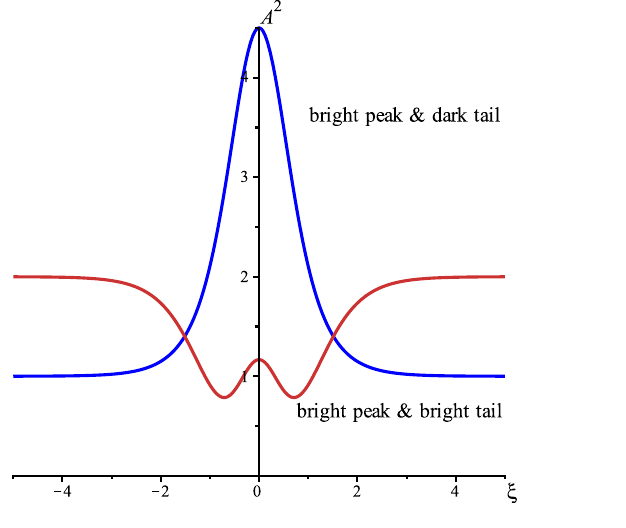}
\includegraphics[trim=2cm 17cm 9cm 2cm,clip,width=0.48\textwidth]{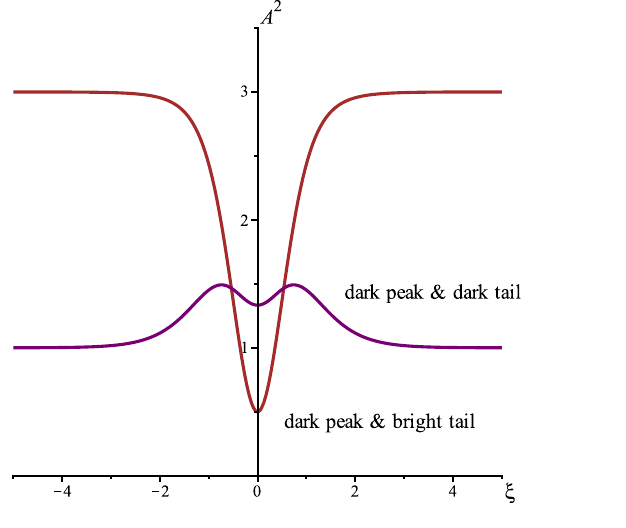}
\caption{Bright/dark peaks and tails for $\H(\z)$}
\label{fig:A.brightdark} 
\end{figure}

The amplitude $\H(\z)$ has the form of a solitary wave 
which is the square-root of a linear combination of
$\V(\z)$, $\V(\z)^2$, $\V(\z)^{p+1}$ terms plus a constant term. 
Note that the resulting profile for $\H(\z)$ can contain up to $2p+1$ local peaks 
on a tail (background) $\H_\infty= \H(\pm\infty)$,
with the central peak at $\z=0$ having the amplitude $\H_0=\H(0)$,
as given by expressions \eqref{Hsq}. 
This central peak will be \emph{bright/dark} if $\H''(0)\gtrless 0$ respectively, 
while the tail will be called \emph{bright/dark} if $\H_\infty \gtrless \H_0$. 
These features are illustrated in Fig.~\ref{fig:A.brightdark}. 

Side peaks occur in symmetrical pairs. 
Their location and height can be found straightforwardly by considering
\begin{equation}
B(z) =  \H\big(\tfrac{1}{\sqrt{a}}\arccosh(1/\sqrt{z})\big)^2 
= g_1 + g_2 (b+s_3h z)  + g_3 (b+s_3h z)^2 - s_1 s_2\tfrac{1}{p+1} (b+s_3h z)^{p+1},
\end{equation}
which is polynomial of degree $p+1$ in $z = \sech(\sqrt{a}\,\z)^2$ with
$0<z\leq 1$. 
In this interval, each non-zero root $z=z^*$ of 
\begin{equation}\label{DHsq.eqn}
B'(z) =  s_3 h( g_2  + 2 g_3 (b+s_3h z) - s_1 s_2 (b+s_3h z)^{p}) =0
\end{equation} 
determines that a pair of side peaks exist at 
$\z^* = \pm\tfrac{1}{\sqrt{a}}\arccosh(1/\sqrt{z^*})\big)^2$. 
These peaks will be bright/dark if $H''(z^*)\gtrless 0$ respectively. 
The case of one pair of side peaks is illustrated in Fig.~\ref{fig:A.brightdark} . 

The phase $\Y(\z)$ is given by a linear term in $\z$ 
plus a quadrature term that is odd in $\z$ with a coefficient $\pm\sqrt{\mu}$. 
In the case when this coefficient vanishes, the phase is thereby linear in $\z$ 
\begin{equation}\label{Psi.muis0}
\Y(\z)  = \frac{c}{2} \z + \phi, 
\quad\text{ if } \mu=0 . 
\end{equation}

In the general case, $\mu >0$, 
the quadrature term in the phase can be expressed more conveniently 
as the integral of a rational function, based on the identities 
$\sech^2(\arctanh(\zeta))=1-\zeta^2$ and $\arctanh'(\zeta)=1/(1-\zeta^2)$: 
\begin{equation}\label{AU.rel}
\int_0^\z\frac{d\z}{\H(\z)^2} 
=  \frac{1}{\sqrt{a}}\int_0^{\tanh(\sqrt{a}\,\z)} \frac{d\zeta}{(1-\zeta^2)B(1-\zeta^2)} , 
\end{equation}
where the degree of $B(1-\zeta^2)$ as a polynomial in $\zeta^2$ is $p+1$. 
%B(\zeta) =  \H\big(\tfrac{1}{\sqrt{a}}\arctanh(\zeta)\big)^2 
For $\z\to \pm\infty$, this quadrature has the asymptotic behaviour
\begin{equation}
\int_0^\z\frac{d\z}{\H(\z)^2} \sim \frac{1}{\H_\infty^2}\, \z +\sgn(\z)\int_0^\infty \Big(\frac{1}{\H(\z)^2}- \frac{1}{\H_\infty^2}\Big)d\z , 
\end{equation}
where $\H_\infty\neq 0$ due to condition \eqref{Hsq.conds}.
This motivates a decomposition of $\Y(\z)$ into linear and nonlinear terms: 
\begin{equation}\label{Psi.muisnot0}
\Y(\z) = \Y_\lin(\z) +\Y_\nonlin(\z) +\phi
\quad\text{ if } \mu>0 , 
\end{equation}
where
\begin{equation}\label{Y.sgn}
\begin{aligned}
\Y_\lin(\z) = & 
\Big(\pm\frac{\sqrt{\mu}}{A_\infty^2} + \frac{c}{2}\Big)\z,
\\
\Y_\nonlin(\z) =&
\frac{\pm\sqrt{\mu}}{\sqrt{a}}\int_0^{\tanh(\sqrt{a}\,\z)}  \Big(\frac{1}{B(1-\zeta^2)} - \frac{1}{A_\infty^2}\Big) \frac{d\zeta}{(1-\zeta^2)} .
\end{aligned}
\end{equation}
Evaluation of $\Y_\nonlin(\z)$ will yield a linear combination of 
arctan and arctanh (or ln) functions in terms of $\zeta=\tanh(\sqrt{a}\,\z)$. 
Note that its graph will be \emph{kink} shaped, since
$\Y_\nonlin(\infty)=-\Y_\nonlin(-\infty)$ and $\Y_\nonlin(0)=0$. 
The sign of the kink is determined by the slope of $\Y_\nonlin(\z)$ at $\z=0$:
\begin{equation}
\sgn(\Y_\nonlin(\infty)) = \sgn(\Y_\nonlin'(0)) = \pm \sgn(A_\infty -A_0)
\end{equation}
using condition \eqref{Hsq.conds},
where ``$\pm$'' refers to the sign in expressions \eqref{Y.sgn}.
Consequently, the asymptotic behaviour of the phase is
\begin{equation} 
\Y(\z) \sim \phi+ \Y_\lin(\z) \pm\Y_\nonlin(\infty) \text{ as }\z\to \pm\infty
\end{equation}
in terms of the asymptote $\Y_\nonlin(\infty)\neq 0$ of the kink. 

All of these features, 
in addition to the sign of the wave speed $c$ and the gKdV-LS coupling $k$, 
are summarized for each nonlinearity power $p=1,2,3,4$ 
in the next section 
(cf Tables~\ref{table:pis1-i.summary} to ~\ref{table:pis4-i.summary}).

\subsection{Features of solitary wave solutions}
\hfill\\

\begin{table}[H]
\hbox{\hspace{-0.75in}
\begin{tabular}{|l|c||c|c|c|c|c|c|c|c|}
\hline
free 
& no. of
& coupling 
& speed
& gKdV 
& 
& LS 
& side
& freq.
& phase 
\\
& fam.
& 
& 
& peak %$U(0)$
& backgrnd. 
& tail \& peak 
& peaks
& parm. 
& kink 
\\
\hline
\hline
$k$, $b$, $c$, $\varpi$
% Maple:pis1-case1
& $1$
& $k>1$
& $c<c_\crit$
& bright
& $b\gtrless0$, $b=0$
& bright \& bright
& $1$
& $\varpi\gtrless0$, $\varpi=0$
& $\mu\geq0$
\\
& 
& 
& $c>c_\crit$
& bright
& $b\gtrless0$, $b=0$
& bright \& dark
& $0$
& $\varpi\gtrless0$, $\varpi=0$
& $\mu>0$
\\
& 
& $0<k<1$
& $c>c_\crit$
& bright
& $b\gtrless0$, $b=0$
& bright \& bright
& $1$
& $\varpi\gtrless0$, $\varpi=0$
& $\mu\geq0$
\\
& 
& 
& $c<c_\crit$
& bright
& $b\gtrless0$, $b=0$
& bright \& dark
& $0$
& $\varpi\gtrless0$, $\varpi=0$
& $\mu>0$
\\
& 
& $k<0$
& $c<c_\crit$
& dark
& $b\gtrless0$, $b=0$
& bright \& bright
& $1$
& $\varpi\gtrless0$, $\varpi=0$
& $\mu\geq0$
\\
& 
& 
& $c>c_\crit$
& dark
& $b\gtrless0$, $b=0$
& bright \& dark
& $0$
& $\varpi\gtrless0$, $\varpi=0$
& $\mu>0$
\\
\hline
$k$, $b$, $c$
% Maple:pis1-case1-subcase2, pis1-case1-subcase3
& $2$
& $k>0$
& $c\gtrless0$
& bright
& $b\gtrless0$, $b=0$
& $0$ \& bright
& $0$
& $\varpi\gtrless0$
& $\mu=0$
\\
& 
& 
& 
& 
& 
& $0$ \& dark
& $1$
& $\varpi\gtrless0$
& $\mu=0$
\\
& 
& $k<0$
& $c\gtrless0$
& dark
& $b\gtrless0$, $b=0$
& $0$ \& bright
& $0$
& $\varpi\gtrless0$
& $\mu=0$
\\
& 
& 
& 
& 
& 
& $0$ \& dark
& $1$
& $\varpi\gtrless0$
& $\mu=0$
\\
\hline
\hline
\end{tabular}
}
\caption{$p=1$: features of solitary wave family (1-i), (1-i.1)
\newline
$c_\crit = 4(1-k) b + (3-4k)\varpi/k$
\newline\vbox{\vskip 0.0in}}
\label{table:pis1-i.summary}
\end{table}

\begin{table}[H]
\hbox{\hspace{-0.7in}
\begin{tabular}{|l|c||c|c|c|c|c|c|c|c|}
\hline
free 
& no. of 
& coupling 
& speed
& gKdV 
&  
& LS tail
& side 
& freq.
& phase 
\\
& fam. 
& 
& 
& peak %$U(0)$
& backgrnd. 
& \& peak 
& peaks
& parm.
& kink 
\\
\hline
\hline
$b$, $c$, $\varpi$, $a$, $s_2$
% Maple:pis1-case3
& $1$
& $k=1/6$
& $c\gtrless0$
& bright
& $b\gtrless0$, $b=0$
& bright \& dark
& $0$
& $\varpi\gtrless 0$, $\varpi=0$
& $\mu\geq0$
\\
\hline
$b$, $c$, $\varpi$, $s_2$
% Maple:pis1-case3-C2sqis0-subcase2
& $1$
& $k=1/6$
& $c\gtrless0$
& bright
& $b\gtrless0$, $b=0$
& $0$ \& bright
& $0$
& $\varpi\gtrless 0$, $\varpi=0$
& $\mu=0$
\\
\hline
\hline
\end{tabular}
}
\caption{$p=1$: features of solitary wave family (1-ii), (1-ii.1)
\newline\vbox{\vskip 0.0in}}
\label{table:pis1-ii.summary}
\end{table}

\begin{table}[H]
\hbox{\hspace{-0.3in}
\begin{tabular}{|l|c||c|c|c|c|c|c|c|c|}
\hline
free 
& no. of 
& coupling 
& speed
& gKdV 
& 
& LS tail
& side 
& freq.
& phase 
\\
& fam.
& 
& 
& peak %$U(0)$
& backgrnd. 
& \& peak 
& peaks
& parm. 
& kink 
\\
\hline
\hline
$b$, $c$, $a$
% Maple:pis1-case4
& $1$
& $k=1$
& $c\gtrless 0$
& bright
& $b<b_\crit$
& bright \& bright
& $1$
& $\varpi\gtrless0$
& $\mu>0$
\\
&
& 
& $c\gtrless 0$
& bright
& $b>b_\crit$
& bright \& dark
& $0$
& $\varpi\gtrless0$
& $\mu>0$
\\
\hline
$b$, $c$
% Maple:pis1-case4-C2sqis0-subcase2, pis1-case4-C2sqis0-subcase3
& $2$
& $k=1$
& $c\gtrless0$
& bright
& $b\gtrless0$, $b=0$
& $0$ \& bright
& $0$
& $\varpi\gtrless0$
& $\mu=0$
\\
& 
& 
& $c\gtrless0$
& bright
& $b\gtrless0$, $b=0$
& $0$ \& dark
& $1$
& $\varpi\gtrless0$
& $\mu=0$
\\
\hline
$b$, $a$
% Maple:pis1-case4-C2sqis0-subcase1
& $1$
& $k=1$
& $c\gtrless0$
& bright
& $b\gtrless0$
& bright \& bright
& $1$
& $\varpi\gtrless0$
& $\mu=0$
\\
\hline
\hline
\end{tabular}
}
\caption{$p=1$: features of solitary wave family (1-iii), (1-iii.1), (1-iii.2)
\newline
$b_\crit = c- 2a$ 
\newline\vbox{\vskip 0.25in}}
\label{table:pis1-iii.summary}
\end{table}

\begin{table}[H]
\hbox{\hspace{-0.5in}
\begin{tabular}{|l|c||c|c|c|c|c|c|c|c|}
\hline
free 
& no. of 
& coupling 
& speed
& gKdV 
& 
& LS tail
& side 
& freq.
& phase 
\\
& families 
& 
& 
& peak %$U(0)$
& backgrnd. 
& \& peak 
& peaks
& parm. 
& kink 
\\
\hline
\hline
$k$, $b$, $c$
% Maple:pis2-case1
& $1$
& $k>0$
& $c\gtrless 0$
& bright
& $b> b_\crit$ 
& bright \& dark
& $0$ 
& $\varpi\gtrless 0$, $\varpi=0$
& $\mu> 0$
\\
& 
& 
& 
& 
& $b= b_\crit$ 
& bright \& dark
& $1$ 
& $\varpi\gtrless 0$, $\varpi=0$
& $\mu> 0$
\\
& 
& 
& 
& 
& $b< b_\crit$ 
& bright \& dark
& $2$ 
& $\varpi\gtrless 0$, $\varpi=0$
& $\mu> 0$
\\
& 
& $k<0$
& $c\gtrless 0$
& dark
& $0<b<b_\crit$ %positive
& bright \& dark 
& $0$ %none
& $\varpi>0$
& $\mu>0$
\\
& 
& 
& 
& 
& $b=b_\crit$ %positive
& bright \& dark 
& $1$ %none
& $\varpi>0$
& $\mu>0$
\\
& 
& 
& 
& 
& $b>b_\crit$ %positive
& bright \& dark 
& $2$ %none
& $\varpi>0$
& $\mu>0$
\\
\hline
$k$, $b$
% Maple:pis2-case1-C2sqis0-subcase2
& $3$
& $k>0$
& $c\gtrless0$
& bright
& $b\gtrless0$, $b=0$
&  bright \& dark
& $2$
& $\varpi\gtrless0$, $\varpi=0$
& $\mu=0$
\\
& 
& 
& $c>0$
& bright
& $b\gtrless 0$, $b=0$
& $0$ \& dark
& $1$
& $\varpi<0$
& $\mu=0$
\\
& 
& 
& $c<0$
& bright
& $b<0$
&  $0$ \& bright
& $2$
& $\varpi>0$
& $\mu=0$
\\
& 
& $k<0$
& $c>0$
& dark 
& $b\gtrless 0$, $b=0$
&  $0$ \& bright 
& $2$
& $\varpi\gtrless0$, $\varpi=0$
& $\mu=0$
\\
& 
& 
& $c<0$
& dark 
& $b>0$
&  bright \& dark 
& $2$
& $\varpi>0$
& $\mu=0$
\\
&
& 
& $c<0$
& dark
& $b>0$
& $0$ \& dark
& $1$
& $\varpi\gtrless0$, $\varpi=0$
& $\mu=0$
\\
\hline
$k$, $c$
% Maple:pis2-C2sqis0-subcase1
& $1$
& $k>0$
& $c\gtrless0$
& bright
& $b<0$
&  $0$ \& bright
& $0$
& $\varpi\gtrless 0$, $\varpi=0$
& $\mu=0$
\\
& 
& $k<0$
& $c>0$
& dark
& $b<0$
&  $0$ \& bright
& $0$
& $\varpi\gtrless 0$, $\varpi=0$
& $\mu=0$
\\
\hline
\hline
\end{tabular}
}
\caption{$p=2$: features of solitary wave family (2-i), (2-i.1), (2-i.2)
\newline
$b_\crit = 2c/|k|$
\newline\vbox{\vskip 0.0in}}
\label{table:pis2-i.summary}
\end{table}

\begin{table}[H]
\hbox{\hspace{-0.2in}
\begin{tabular}{|l|c||c|c|c|c|c|c|c|c|}
\hline
free 
& no. of 
& coupling 
& speed
& gKdV 
& 
& LS tail
& side 
& freq.
& phase 
\\
& families 
& 
& 
& peak %$U(0)$
& backgrnd. 
& \& peak 
& peaks
& parm. 
& kink 
\\
\hline
\hline
$b$, $\varpi$
% Maple:pis2-case2
& $1$
& $k<0$
& $c< 0$
& dark
& $b>0$ %positive
& bright \& dark
& $0$ 
& $\varpi>\varpi_\side$
& $\mu\geq0$
\\
& 
& 
& 
& 
& 
& 
& $1$ 
& $\varpi=\varpi_\side$
& $\mu\geq0$
\\
& 
& 
& 
& 
& 
& 
& $2$ 
& $\varpi<\varpi_\side$
& $\mu\geq0$
\\
\hline
$b$
% Maple:pis2-case2-C2sqis0
& $2$
& $k>0$
& $c< 0$
& bright
& $b<0$ %negative
& $0$ \& bright
& $2$ 
& $\varpi>0$
& $\mu=0$
\\
& 
& $k<0$
& $c< 0$
& dark
& $b>0$ %positive
& $0$ \& dark
& $1$ 
& $\varpi>0$
& $\mu=0$
\\
\hline
\hline
\end{tabular}
}
\caption{$p=2$: features of solitary wave family (2-ii), (2-ii.1)
\newline
$\varpi_\side = 17b^2/16$
\newline\vbox{\vskip 0.0in}}
\label{table:pis2-ii.summary}
\end{table}

\begin{table}[H]
\hbox{\hspace{-0.25in}
\begin{tabular}{|l|c||c|c|c|c|c|c|c|c|}
\hline
free 
& no. of 
& coupling 
& speed
& gKdV 
& 
& LS tail
& side 
& freq. 
& phase 
\\
& families 
& 
& 
& peak %$U(0)$
& backgrnd. 
& \& peak 
& peaks
& parm. 
& kink 
\\
\hline
\hline
$k,b$
% Maple:pis3
& $1$
& $k>0$
& $c\gtrless 0$
& bright
& $b<b_\side<0$ %negative
& bright \& bright 
& $1$ 
& $\varpi>0$
& $\mu\geq 0$
\\
& 
& 
& 
& 
& $b=b_\side<0$ %negative
& bright \& bright 
& $2$ 
& $\varpi>0$
& $\mu\geq 0$
\\
& 
& 
& 
& 
& $0>b>b_\side$ %negative
& bright \& bright 
& $3$ 
& $\varpi>0$
& $\mu\geq 0$
\\
& 
& $k<0$
& $c<0$
& dark
& $b>b_\crit>0$ %positive
& bright \& bright
& $1$ %none
& $\varpi>0$
& $\mu>0$
\\
& 
& 
& 
& 
& $0<b<b_\crit$ %positive
& bright \& dark 
& $0$ %none
& $\varpi>0$
& $\mu> 0$
\\
\hline
$k$
% Maple:pis3-C2sqis0
& $3$
& $k>0$
& $c< 0$
& bright
& $b<0$ %negative
& $0$ \& bright 
& $2$ 
& $\varpi>0$
& $\mu=0$
\\
& 
& 
& 
& 
& 
& $0$ \& dark 
& $1$
& $\varpi>0$
& $\mu=0$
\\
& 
& 
& 
& 
& 
& $0$ \& dark 
& $3$
& $\varpi>0$
& $\mu=0$
\\
\hline
\hline
\end{tabular}
}
\caption{$p=3$: features of solitary wave family (3-i), (3-i.1)
\newline
$b_\side$ is the real negative root of $25 b^6 - 125k b^4 + 204k^2 b^2 - 180k^3=0$
\newline
$b_\crit$ is the real positive root of $5b^6 - 131|k| b^4 - 21|k|^2 b^2 - 45|k|^3 =0$
\newline\vbox{\vskip 0.5in}}
\label{table:pis3-i.summary}
\end{table}

\begin{table}[H]
%\hbox{\hspace{-0.4in}
\begin{tabular}{|l|c||c|c|c|c|c|c|c|c|}
\hline
free 
& no. of 
& coupling 
& speed
& gKdV 
& 
& LS tail
& side 
& freq.
& phase 
\\
& families 
& 
& 
& peak %$U(0)$
& backgrnd. 
& \& peak 
& peaks
& parm.
& kink 
\\
\hline
\hline
$k$
% Maple:pis4
& $1$
& $k>0$
& $c<0$
& bright
& $b<0$ %negative
& bright \& dark 
& $0$ %none
& $\varpi>0$
& $\mu>0$
\\
\hline
\hline
\end{tabular}
%}
\caption{$p=4$: features of solitary wave family (4-i)}
\label{table:pis4-i.summary}
\end{table}

\subsection{Conserved integrals}

All of the solitary wave solutions \eqref{soln}
with expressions \eqref{U.sol}, \eqref{A.sol}, \eqref{Psi.sol} 
have a total mass, charge, momentum, and energy 
given respectively by the four conserved integrals \eqref{M}--\eqref{H}.
The latter two integrals \eqref{P}--\eqref{H} can be simplified 
by use of the ODEs \eqref{V'.eqn} for $\V(\z)$ and \eqref{Y'.eqn} for $\Y(z)$. 

Each integral will be finite if $u$ has a zero background and $|\psi|$ has a zero tail, 
namely $b=0$ and $g_1=0$, 
since then the density goes to $0$ exponentially in $|\z|$. 

If a density otherwise has a non-zero asymptotic tail as $|\z|\to\infty$, 
the conserved integral will diverge. 
Nevertheless, the divergent contribution can be removed by subtraction of 
the asymptotic tail in the density, yielding a regularized integral 
which is conserved and finite. 

The regularized integrals for mass, charge, momentum, and energy are given by 
\begin{align}
& \tilde{M}[\V] =\int_{-\infty}^{\infty} (\V-b)\, d\z, 
\label{regM}\\
& \tilde{J}[\H] =\int_{-\infty}^{\infty}\tfrac{1}{2} (\H^2 -\H_0^2)\, d\z, 
\label{regJ}\\
& \tilde{P}[\V,\H,\Y] =\int_{-\infty}^{\infty}\big( 
\tfrac{1}{2} (\V^2 -b^2) -s_2\tfrac{c}{2k} (\H^2 -\H_0^2) 
\big)\, d\z, 
\label{regP}\\
&\begin{aligned} 
\tilde{H}[\V,\H,\Y] =& \int_{-\infty}^{\infty}\big( 
\tfrac{1}{2} \V'{}^2 +s_2\tfrac{1}{k} \H'{}^2 
- s_1\tfrac{1}{(p+1)(p+2)}(\V^{p+2} -b^{p+2}) 
\\&\qquad
+s_2\tfrac{c^2}{4k} (\H^2 -\H_0^2)  -s_2 (\V\H^2 -b\H_0^2)
\big)\, d\z
\\
=& \int_{-\infty}^{\infty}\big( 
\tfrac{1}{2}(\V^2 -b^2) + C_1 (\V -b) - s_1\tfrac{2}{(p+1)(p+2)}(\V^{p+2} -b^{p+2}) 
\\&\qquad
-s_2\tfrac{\omega}{k} (\H^2 -\H_0^2)  -s_2 2(\V\H^2 -b\H_0^2)
\big)\, d\z , 
\end{aligned}
\label{regH}
\end{align}
which are finite for all solitary wave solutions \eqref{U.sol}, \eqref{A.sol}, \eqref{Psi.sol}.

It is straightforward to evaluate the mass integral \eqref{regM}:
\begin{equation}\label{regM.eval}
\tilde{M}[\V] =\int_{-\infty}^{\infty} 
s_3 h\, \sech^2(\sqrt{a}\, \z) d\z
= s_3 \frac{2 h}{\sqrt{a}} .
\end{equation}
This is the same general expression as the mass of the mKdV solitary wave
and depends on $p$ only implicitly through $h$ when it is evaluated for 
a specific solution family. 

It is similarly straightforward to evaluate the charge integral \eqref{regJ}
for each power $p=1,2,3,4$,
where the density is given by 
$\frac{1}{2} g_3 h^2\, \sech^4(\sqrt{a}\, \z) 
+ s_3\frac{1}{2} (2 b g_3 +g_2)h\, \sech^2(\sqrt{a}\, \z) 
+ s_1 s_2\frac{1}{2(p+1)} \big( b^{p+1} - \big( s_3 h\, \sech^2(\sqrt{a}\, \z) +b\big)^{p+1} \big)$. 
Alternatively, a single expression that holds for all $p$ can be obtained 
by applying the following integration steps to the third term in the density. 
Make a binomial expansion of this term 
and change the integration variable via $\sech(\sqrt{a}\,\z) = 2y/(1+y^2)$. 
Since the density is even in $\z$, the charge integral can be expressed as 
twice the expanded density integrated over $y$ from $0$ to $1$.
Each term is given by an associated Legendre function and the sum of all the terms
can be shown to yield a generalized hypergeometric function ${}_3F_2$
\cite{AbrSte}. 
When combined with the integral of the first and second terms in the density, 
this gives a simple general expression 
\begin{equation}\label{regJ.eval}
\tilde{J}[\V] = s_3\frac{h}{\sqrt{a}}\Big( 
2 g_3(b + s_3 h/3) + g_2 + s_1 s_2 b^p\, {}_3F_2\big( [1,1,-p], [3/2,2],{-s_3}h/b\big) 
\Big),
\quad
p=1,2,3,4 .
\end{equation}
This expression involves $p$ explicitly as well as implicitly through $b,h,g_2,g_3$ 
when it is evaluated for a specific solution family. 
The generalized hypergeometric function ${}_3F_2\big( [1,1,-p], [3/2,2],{-s_3}h/b\big)$ 
is a polynomial of degree $p$ in $h/b$. 

The same steps can be used to evaluate the momentum and energy integrals,
which yields the respective expressions
\begin{gather}
\begin{aligned}
\tilde{P}[\V,\H,\Y] = s_3\frac{h}{k\sqrt{a}}\Big( &
2(k - s_2 c g_3)(b+ s_3 h/3) - s_2 c g_2
\\&\quad
+ s_1 c b^p\, {}_3F_2\big( [1,1,-p], [3/2,2],{-s_3}h/b\big) 
\Big),
\quad
p=1,2,3,4 ; 
\end{aligned}
\label{regP.eval}
\\
\begin{aligned}
\tilde{E}[\V,\H,\Y] = s_3\frac{h}{\sqrt{a}}\Big( &
s_2 4\big( 
g_3 (b+ s_3 h/3)^2 
- (g_3(b+ s_3 h/3) + g_2/2)(4(b+ s_3 h/3) + \omega/k)
\\&\qquad
- g_3 h^2/5 -g_1
\big)
+ 2c (b+ s_3 h/3) + 2C_1 
\\&\quad
+ s_1 2 b^p\big( 2b \, {}_3F_2\big( [1,1,-p-1], [3/2,2],{-s_3}h/b\big) 
\\&\qquad\qquad\quad
+ \omega/k\, {}_3F_2\big( [1,1,-p], [3/2,2],{-s_3}h/b\big) \big)
\Big),
\quad
p=1,2,3,4 . 
\end{aligned}
\label{regE.eval}
\end{gather}

\begin{prop}\label{prop:reg.conservedintegrals}
Solitary wave solutions \eqref{U.sol}, \eqref{A.sol}, \eqref{Psi.sol} 
have regularized mass \eqref{regM.eval}, charge \eqref{regJ.eval}, momentum \eqref{regP.eval}, and energy \eqref{regE.eval},
all of which are finite. 
\end{prop}

\subsection{Remarks on stability}

A basic notion of stability for a solution $(\V(\z),\H(\z),\Y(z))$ is linear (spectral) stability 
under perturbations. 
For a parametric family of solutions, a stronger notion is orbital stability. 
Both notions are related to convexity of the integral quantity \eqref{W} 
in the variational principle \eqref{W.ELeqns} for the ODE system \eqref{U.A.Phi.sys} 
that determines solutions. 
In the case of solutions that have a non-zero background,  
a regularized version of this integral quantity can be used,
\begin{equation}\label{W.reg}
\widetilde{W}[\V,\H,\Y] = \tilde{H}[\V,\H,\Y] + c\tilde{P}[\V,\H,\Y] -s_2 (2\omega/k) \tilde{J}[\H] + C_1\tilde{M}[\V] , 
\end{equation}
where the conserved integrals \eqref{M}--\eqref{H} are replaced by 
their regularized counterparts \eqref{regM}--\eqref{regH}, 
which does not change the Euler-Lagrange ODEs. 

Convexity is determined by the eigenvalues of the Hessian operator of 
the regularized integral quantity \eqref{W.reg}. 
After reverting from amplitude and phase variables to the LS variable \eqref{amplitudephase},
the Hessian is given by matrix operator 
\begin{equation}
\begin{aligned}
& \partial^2 \widetilde{W}[\V,\Psi,\bar\Psi;p] = 
\\
& 
- \begin{pmatrix}
\dfrac{d^2}{d\xi^2} -c  +s_1 \V^p
& 
s_2\bar\Psi
& 
s_2\Psi
\\
s_2\Psi
& 
(s_2/k)\Big( \dfrac{d^2}{d\xi^2} - i c \dfrac{d}{d\xi} +\omega \Big) +s_2 \V
& 
0 
\\
s_2\bar\Psi
& 
0 
& 
(s_2/k)\Big( \dfrac{d^2}{d\xi^2} + i c \dfrac{d}{d\xi} +\omega \Big) +s_2 \V
\end{pmatrix} . 
\end{aligned}
\end{equation}
This operator is formally self-adjoint with respect to a joint 
$L^2$ inner product for the $\V$-component 
and a hermitian $L^2$ inner product for the $\Psi,\bar\Psi$-components. 
It can be brought to a diagonal form by the change of variables
\begin{equation}\label{diag.vars}
\Psi = s_2 (k/2) e^{i\frac{c}{2}\xi} (\Theta_1 + i\Theta_2),
\quad
\bar\Theta_1=\Theta_1,
\quad
\bar\Theta_2=\Theta_2, 
\end{equation}
yielding 
\begin{equation}\label{Lop}
\Lop = \Iop \dfrac{d^2}{d\xi^2} +\Vop
\end{equation}
with 
\begin{equation}
\Iop = \diag\big(1,s_2 k/2,s_2 k/2\big), 
\quad
\Vop = 
\begin{pmatrix}
-c  +s_1 \V^p
& 
s_2 (k^2/2)\Theta_1
& 
s_2 (k^2/2)\Theta_2
\\
s_2 (k^2/2)\Theta_1
& 
s_2 (k/2)(\varpi  + k \V)
& 
0 
\\
s_2 (k^2/2)\Theta_2
& 
0
&
s_2 (k/2)(\varpi  + k \V) 
\end{pmatrix} , 
\end{equation}
where $\varpi$ is expression \eqref{w.rel}. 
This is a linear self-adjoint matrix operator with respect to a real $L^2$ inner product
for the real components $(U,\Theta_1,\Theta_2)$. 

From Proposition~\ref{prop:ODE.pointsymms},
there are three infinitesimal symmetries \eqref{ODE.symms}--\eqref{ODE.symms.pis1}
of the Euler-Lagrange ODEs \eqref{U.A.Phi.sys}. 
A standard argument shows that the translation and phase rotation symmetries 
will each give rise to a zero eigenvalue of the operator \eqref{Lop}. 
The key question for determining stability is whether 
this operator has a negative eigenvalue \cite{GeyPel-book}. 
This will be addressed in subsequent work.

\section{Concluding remarks}\label{sec:conclude}

The new method developed in the present work 
has yielded 22 families of exact solitary wave solutions \eqref{inv.soln} 
for the coupled nonlinear gKdV-LS system \eqref{u.eqn.phys}--\eqref{psi.eqn.phys}
with nonlinearity powers $p=1,2,3,4$. 
No solutions for $p>1$ were known previously. 

For $p=1$, 
four of the solution families contain the solitary waves that were first found 
in \Ref{NisHojMimIke,CisPel2017}. 
The former one arose in the context of nonlinear plasma waves 
and coincides with the first of the two families 1-i.1 (i.e.\ dark LS peak). 
The latter ones come from nonlinear waves in anharmonic lattices
and were named bright/dark Davydov solitons of the 1st and 2nd kind: 
they are respectively given by 
the case $s_2=-1$ of families 1-ii with $\mu\neq0$ and 1-ii.1; 
and the case $k=1/6$ of family 1-i with $\mu=0$ and $c<c_\crit$
and the case $k=1/6$ of the second family 1-i.1 (i.e.\ bright LS peak). 
These particular solitary waves are characterized by respective relations 
$|\psi|^2 \propto$ $(u-b)(u- u_0)$, $u-b$, $(u-u_0)^2$, 
for some constant $u_0$,
which allows them to be derived by basic ODE methods. 
The corresponding relations for the travelling wave ODE system \eqref{ODE.sys} 
have the form $\H^2 \propto$ $(\V- b)(\V-\V_0)$, $\V-b$, $(\V-\V_0)^2$ 
in terms of the variables \eqref{amplitudephase}. 

The additional five solution families for $p=1$ describe new solitary waves. 
Physical properties and applications, 
as well as stability analysis based on the variational principle \eqref{W.ELeqns}, 
for all of the new solution families for $p=1,2,3,4$ will be discussed elsewhere. 

The main results obtained here illustrate the power of 
modern symmetry analysis methods combined with advanced symbolic computation packages 
for the study of exact solitary wave solutions of coupled nonlinear ODE systems. 

As further applications, 
one direction would be a generalization of the gKdV-LS system to include 
a self-interaction $\f |\psi|^q \psi$ in equation \eqref{psi.eqn.phys} for the Schrodinger field $\psi$, 
where $q>0$ is a nonlinearity power. 
Only a few exact solutions are known for this nonlinear coupled system \cite{LabEbaZerBis}. 
Another direction would be to consider a coupled Boussinesq-LS system \cite{GaiChrMin}
and its $p$-power generalizations,
for which exact solitary wave solutions are of physical interest. 

A more general direction would be to extend the present method to the study of
nonlinearly generalized H\'enon-Heiles ODE systems,
which arise in wide variety of physical and applied mathematical problems.

\appendix
\section{Computational remarks}\label{sec:computation}

The computation leading to the results summarized in Table~\ref{table:soln.cases}
will be outlined here. 

As explained in Section~\ref{sec:classify}, 
the computation involves solving a polynomial equation obtained by 
substitution of the general polynomial ansatz \eqref{G.ansatz}
into the ODE \eqref{G.eqn} for $G(\V)$. 
This polynomial equation has 436 terms, 
containing 16 powers $0, 1, 2, 3, 4, 5, p, p+1, p+2, p+3, p+4, 2p, 2p+1, 2p+2, 2p+3$.
There are 14 unknowns, consisting of 
$p$, $k$, $b$, $h$, $a$, $c$, $\varpi$, $\mu$, $C_1$, $\tilde C_3$, $E$,
and $s_1,s_2,s_3$, 
which are required to obey 8 conditions, 
$p\neq 0,-1$, $k\neq 0$, $h\neq 0$, $a\neq 0$, $c\neq0$, 
and $s_1{}^2=s_2{}^2=s_3{}^2=1$. 
These conditions together with the polynomial equation 
constitute the algebraic system to be solved. 
This is a non-polynomial algebraic problem, because $p$ appears in exponents 
and because products of the unknowns (including $p$) appear in the coefficients of some terms. 

The computer algebra package Crack \cite{Wol}, written in REDUCE, 
is able to solve nonlinear problems (algebraic as well as differential).  
Specifically, it is able to split equations with respect to independent variables, including when a variable appears as an exponent.
It is also able to make case distinctions as needed. 
In an automatic run, all splitting with respect to the independent variables 
has a high priority. 

The system here has a single independent variable, $\V$. 
Crack splits the system with respect to $\V$ if all exponents are known to be different. 
Since $p$ is an unknown that appears in exponents, 
Crack proceeds by systematically taking the difference between two exponents, 
with at least one of the exponents containing $p$, 
and starting a case distinction in which the pivot is given by the expression 
for the difference being zero or non-zero. 
In the non-zero case, this inequality is simply appended to the system. 
In the zero case,
Crack determines $p$ and then substitutes it back into the system. 
If the resulting system either has all exponents of $\V$ being distinct, 
or has no exponents of $\V$ involving $p$, 
then it can be split into a system of algebraic equations that involve only the unknowns. 
Crack is able to solve these equations straightforwardly. 
If an equation factorizes, 
then a further case distinction is made with respect to each factor. 
This process is repeated until all possible case distinctions have been considered, 
and the system has been either fully solved or reduced to irreducible equations 
such as high degree polynomials in the unknowns.  
In the system under consideration, 
every case leads to an algebraic system that is explicitly solved
in the sense that its solution is given by explicit expressions 
together with at most low-degree polynomials whose solution set is non-empty. 

The final result is 56 solutions that can be merged into non-overlapping families 
organized by $p$ as shown in Table~\ref{table:output.cases}. 

\begin{table}[H]
%\hbox{\hspace{-0.0in}
\begin{tabular}{c|c||l|l}
\hline
nonlinearity
& no. of 
& determined
& free 
\\
$p$
& families &&
\\
\hline
\hline
$4$ 
& $1$
& $b$, $h$, $a$, $\mu$, $\varpi$, $c$; $s_2$
& $k$; $s_1$, $s_3$
\\
\hline
$3$ 
& $1$
& $h$, $a$, $\mu$, $\varpi$, $c$; $s_1$
& $k$, $b$; $s_2$, $s_3$
\\
\hline
$2$ 
& $2$
& $h$, $a$, $\mu$, $\varpi$; $s_2$
& $k$, $b$, $c$; $s_1$, $s_3$
\\
& 
& $b$, $h$, $a$, $\mu$, $c$; $s_2$
& $k$, $\varpi$; $s_1$, $s_3$
\\
\hline
$1$ 
& $3$
& $a$, $h$, $\mu$; $s_1$
& $k$, $b$, $c$, $\varpi$; $s_2$, $s_3$
\\
& 
& $k$, $h$, $\mu$; $s_1$
& $b$, $a$, $c$, $\varpi$; $s_2$, $s_3$
\\
&
& $k$, $h$, $\mu$, $\varpi$; $s_1$
& $b$, $a$, $c$; $s_2$, $s_3$
\\
\hline
\hline
\end{tabular}
%}
\caption{Solution cases}
\label{table:output.cases}
\end{table}

\section{Solution families}\label{sec:output}

Firstly, the output for the non-overlapping families of solutions listed in Table~\ref{table:output.cases} 
will be summarized. 
Secondly, for each solution family, 
the conditions \eqref{Hsq.conds} and \eqref{V.conds} will be analyzed 
to derive a minimal set of inequalities on the free parameters. 

In the special case $\mu=0$ and $\H(0)=\H(\pm\infty)=0$ 
when the necessary conditions \eqref{Hsq.conds} are not sufficient, 
the following stronger conditions will be used:
\begin{equation}\label{Hsq.muis0.conds}
\begin{gathered}
s_3 \big( g_2 +2 g_3 (b+s_3 h)  - s_1 s_2 (b+s_3 h)^p \big) \leq 0,
\\
s_3 \big( g_2 +2 g_3 b - s_1 s_2 b^p \big) \geq 0 .
\end{gathered}
\end{equation}
These conditions are derived straightforwardly 
from an asymptotic expansion of expression \eqref{A.sol} 
for $\H(\z)^2$ with $\z\sim 0$ and $\z\sim \pm\infty$, respectively,
applied to the full condition \eqref{Hsq.nonneg}.
The expansion uses 
$U(\z) -U(0) \sim -s_3 h \z^2$ as $\z\sim 0$
where $U(0) = b + s_3 h$, 
and 
$U(\z) - b \sim 2 s_3 h e^{-\sqrt{a}|\z|}$ as $\z\sim \pm\infty$
from expression \eqref{U.sol}.

{\bf Solution families for $\boldsymbol{p=1}$}:

The \underline{first family} has free parameters 
$b$, $c$, $\varpi$, $k\neq0,1$.
The determined parameters and signs are given by
%maple pis1-case1
\begin{subequations}\label{params.pis1.fam1}
\begin{gather}
s_1=1,
\quad
a = -\frac{kc+(2k-1)\varpi}{4(k-1)} - \frac{kb}{2}, 
\quad
h = s_3\frac{6a}{k}, 
%h = -s_3\Big( \frac{3(kc+(2k-1)\varpi)}{2k(k-1)} + 3b \Big),
\\
g_1 = s_2 \frac{(2k - 1)(4k^2(c + \varpi)^2 +3(kc + \varpi)^2)}{8k^2(k - 1)^2} -\frac{6 a^2}{k^2},
\\
g_2 = s_2\frac{(2k - 1)(c + \varpi)}{k - 1},
\quad
g_3 = s_2 k ,
\\
E = b^2\Big( \frac{kb}{3} + 2a \Big), 
\\
\begin{aligned}
\mu = &
\frac{(2k - 1)^2}{4k^2(k - 1)^4} (kb+\varpi) ((k-1)b +c +\varpi)^2 (2(k-1)(kb +\varpi) -(kc +\varpi))^2, 
\end{aligned}
\end{gather}
\end{subequations}
which are subject to the conditions \eqref{Hsq.conds}--\eqref{V.conds}. 

The first two of the three conditions \eqref{V.conds} yield
\begin{equation}\label{signs1.pis1.fam1}
s_3=\sgn(k)
\end{equation}
and
\begin{equation}\label{ineqn1.pis1.fam1}
2(kb +\varpi)  + \frac{kc +\varpi}{k-1} <0 , 
\end{equation}
while the third condition gives
\begin{equation}\label{ineqn2.C2sqnot0.pis1.fam1}
kb +\varpi >0
\quad\text{when $\mu\neq0$},
\end{equation}
and 
\begin{equation}\label{ineqn2.C2sqis0.pis1.fam1}
\begin{gathered}
(2k-1)(kb +\varpi) ( kc +\varpi + (k-1)(kb +\varpi) ) ( kc +\varpi -2(k-1)(kb +\varpi) ) =0
\\
\quad\text{when $\mu=0$}.
\end{gathered}
\end{equation}

In the case $\mu\neq0$, 
conditions \eqref{Hsq.conds.C2sqnot0} combined with the inequalities \eqref{ineqn1.pis1.fam1}--\eqref{ineqn2.C2sqnot0.pis1.fam1}
are found to yield
\begin{equation}
s_2 =\sgn(k-1/2) . 
\end{equation}
The inequalities themselves reduce to 
\begin{equation}\label{ineqns.C2sqnot0.pis1.fam1}
\begin{aligned}
& c> 2(1-k)b -(2k -1)\varpi/k  
\quad\text{ if $0<k<1$} , 
\\
& c< 2(1-k)b -(2k -1)\varpi/k
\quad\text{ if $k<0$ or $k>1$} . 
\end{aligned}
\end{equation}

In the case $\mu=0$, 
condition \eqref{ineqn2.C2sqis0.pis1.fam1} splits into four subcases:
\begin{align}
& \varpi = -kb , 
\label{subcase1.C2sqis0.pis1.fam1}
\\
& \varpi = -(k-1)b -c , 
\label{subcase2.C2sqis0.pis1.fam1}
\\
& \varpi = -\frac{k(2(k-1)b -c)}{2k-3} , 
\label{subcase3.C2sqis0.pis1.fam1}
\end{align}
and $k=1/2$. 
In the latter subcase, $\H(\z)$ turns out to be identically zero, 
which need not be considered further. 

For subcase \eqref{subcase1.C2sqis0.pis1.fam1}, 
the inequality \eqref{ineqn1.pis1.fam1} combined with conditions \eqref{Hsq.conds.C2sqis0} give
\begin{equation}\label{sign.C2sqis0.pis1.fam1}
s_2 = \sgn(k-1/2), 
\quad
k\neq 1/2
\end{equation}
and
\begin{equation}\label{ineqns.subcase1.C2sqis0.pis1.fam1}
\begin{aligned}
& c> b
\quad\text{ if $0<k<1$} , 
\\
& c< b 
\quad\text{ if $k<0$ or $k>1$} . 
\end{aligned}
\end{equation}
Similarly, for subcase \eqref{subcase2.C2sqis0.pis1.fam1}, 
the inequality \eqref{ineqn1.pis1.fam1} combined with conditions \eqref{Hsq.conds.C2sqis0} 
give the sign condition \eqref{sign.C2sqis0.pis1.fam1} 
and 
\begin{equation}\label{ineqns.subcase2.C2sqis0.pis1.fam1}
c> b . 
\end{equation}
In subcase \eqref{subcase3.C2sqis0.pis1.fam1}, 
conditions \eqref{Hsq.conds.C2sqis0} hold as equalities
and therefore the further condition \eqref{Hsq.muis0.conds} is necessary,
This yields
\begin{equation}\label{sign.subcase3.C2sqis0.pis1.fam1}
s_2 = -\sgn(k-1/2), 
\quad
k\neq 1/2 , 
\end{equation}
while the inequality \eqref{ineqn1.pis1.fam1} gives
\begin{equation}\label{ineqns.subcase3.C2sqis0.pis1.fam1}
\begin{aligned}
& c> b
\quad\text{ if $0<k<3/2$}, 
\\
& c< b 
\quad\text{ if $k<0$ or $k>3/2$} .
\end{aligned}
\end{equation}

Note that the subcase 
\eqref{subcase1.C2sqis0.pis1.fam1}
can be merged with the main case
\eqref{ineqn1.pis1.fam1},
since inequalities \eqref{ineqn2.C2sqnot0.pis1.fam1}, \eqref{ineqns.C2sqnot0.pis1.fam1} 
reduce respectively to inequalities \eqref{sign.C2sqis0.pis1.fam1} and \eqref{ineqns.subcase1.C2sqis0.pis1.fam1}
when \eqref{subcase1.C2sqis0.pis1.fam1} holds.

The \underline{second family} has free parameters $b$, $c$, $\varpi$, $a>0$. 
The determined parameters and signs are given by
%maple pis1-case3
\begin{subequations}\label{params.pis1.fam2}
\begin{gather}
s_1=1, 
\quad
k = \frac{1}{6},
\quad
h = s_3 12 a , 
\\
g_1 = s_2 3(b - c + 4a)(b + 4a + 4\varpi), 
\quad
g_2 = -s_2 (b - c + 4a), 
\quad
g_3 = s_2\frac{1}{2},
\\
E = \frac{b^2 (b + 12a)}{6},
\quad
\mu = \frac{2(b - c + 4a)^2 (b + 6(\varpi+a))^2 (b + 6\varpi)}{3} . 
\end{gather}
\end{subequations}

The three conditions \eqref{V.conds} give 
\begin{equation}
s_3=1
\end{equation}
and
\begin{gather}
b + 6\varpi >0
\quad\text{when $\mu\neq0$} , 
\label{ineqn1.C2sqnot0.pis1.fam2}
\\
(b+6\varpi) (b+6(\varpi + a)) (b -c +4a) =0
\quad\text{when $\mu=0$} . 
\label{ineqn1.C2sqis0.pis1.fam2}
\end{gather}

For $\mu\neq0$, 
the two conditions \eqref{Hsq.conds.C2sqnot0} reduce to
$0 < s_2 (b - c + 4a)$ after use of $a>0$. 
This yields 
\begin{equation}
\begin{aligned}
& c < b + 4a \text{ if } s_2 =1 
\\
& c > b + 4a \text{ if } s_2 =-1 
\end{aligned}
\quad\text{when $\mu\neq0$} , 
\end{equation}
which are independent of inequality \eqref{ineqn1.C2sqnot0.pis1.fam2}.

For $\mu=0$, 
condition \eqref{ineqn1.C2sqis0.pis1.fam2} splits into three subcases:
\begin{align}
& \varpi = -\frac{b}{6} , 
\label{subcase1.C2sqis0.pis1.fam2}
\\
& a = -\frac{b}{6} - \varpi >0 , 
\label{subcase2.C2sqis0.pis1.fam2}
\end{align}
and $a = (c-b)/4>0$. 
In the latter subcase, $\H(\z)$ is found to be identically zero, 
which will not be considered further. 

In subcase \eqref{subcase1.C2sqis0.pis1.fam2}, 
the two conditions \eqref{Hsq.conds.C2sqis0} give
\begin{equation}
\begin{aligned}
& c < b + 4a \text{ if } s_2 =1 , 
\\
& c > b + 4a \text{ if } s_2 =-1 .  
\end{aligned}
\end{equation}
In subcase \eqref{subcase2.C2sqis0.pis1.fam2}, 
conditions \eqref{Hsq.conds.C2sqis0} similarly yield
\begin{equation}
\begin{aligned}
& c > \frac{b}{3} -4\varpi \text{ if } s_2 =1 , 
\\
& c < \frac{b}{3} -4\varpi \text{ if } s_2 =-1 . 
\end{aligned}
\end{equation}
There are no further conditions. 

Note that the subcase \eqref{subcase1.C2sqis0.pis1.fam2} with $\mu=0$ 
and the case \eqref{ineqn1.C2sqnot0.pis1.fam2} with $\mu\neq0$
can be merged: 
$\varpi \geq -b/6$ with $\mu\geq 0$.

The \underline{third family} has free parameters $b$, $c$, $a>0$. 
The determined parameters and signs are given by
%maple pis1-case4
\begin{subequations}\label{params.pis1.fam3}
\begin{gather}
s_1=1, 
\quad
k = 1, 
\quad
h = s_3 6 a ,
\\
g_1 = s_2\Big( \frac{7(7b - 8c + 8a)(7b - 2c + 8a)}{4} -\frac{3b(7b -10c)}{8} \Big),
\quad
g_2 = -s_2 (2b -c + 4a),
\quad
g_3 = s_2, 
\\
E = \frac{b^2 (b + 6a)}{3},
\quad
\mu =4(b -c)(b -c +a)^2(b -c + 4a)^2,
\quad
\varpi=-c . 
\end{gather}
\end{subequations}

The three conditions \eqref{V.conds} give 
\begin{equation}
s_3=1
\end{equation}
and
\begin{gather}
b -c  >0 
\quad\text{when $\mu\neq0$} , 
\label{ineqn1.C2sqnot0.pis1.fam3}
\\
(b -c) (b -c + a) (b -c +4a) =0
\quad\text{when $\mu=0$} . 
\label{ineqn1.C2sqis0.pis1.fam3}
\end{gather}

For $\mu\neq0$, 
the two conditions \eqref{Hsq.conds.C2sqnot0} reduce to 
\begin{equation}
s_2 =1 
\end{equation}
due to $a>0$. 

For $\mu=0$, 
condition \eqref{ineqn1.C2sqis0.pis1.fam3} splits into three subcases:
\begin{align}
& c=b , 
\label{subcase1.C2sqis0.pis1.fam3}
\\
& a = \frac{c-b}{4} >0 , 
\label{subcase2.C2sqis0.pis1.fam3}
\\
& a = c-b >0 . 
\label{subcase3.C2sqis0.pis1.fam3}
\end{align}
In both subcases \eqref{subcase1.C2sqis0.pis1.fam3} and \eqref{subcase2.C2sqis0.pis1.fam3},
the two conditions \eqref{Hsq.conds.C2sqis0} reduce to 
\begin{equation}
s_2 =1 . 
\end{equation}
In subcase \eqref{subcase3.C2sqis0.pis1.fam3},
conditions \eqref{Hsq.conds.C2sqis0} hold as equalities, 
and so the stronger condition \eqref{Hsq.muis0.conds} is necessary.
This stronger condition yields
\begin{equation}
s_2 =-1 .  
\end{equation}
There are no further conditions in any of the subcases.

{\bf Solution families for $\boldsymbol{p=2}$}:

The \underline{first family} has free parameters $b$, $c$, $k\neq0$. 
The determined parameters and signs are given by 
%maple pis2-case1
\begin{subequations}\label{params.pis2.fam1}
\begin{gather}
s_2=1 , 
\\
a = \frac{(2c - 4kb + s_1(2b^2 + k^2))k}{16(k -s_1 b)},
\quad
h = s_3 \frac{12a}{k},
\\
g_1 = \frac{k -s_1 2 b}{(k -s_1 b)^2}\Big( c +\frac{3kb}{16} -s_1\frac{5k^2}{16} - s_1\frac{3b^2}{8} \Big)^2
 -\Big(k -s_1\frac{22b}{3}\Big)\Big(\frac{3k}{16} - s_1\frac{b}{8}\Big)^2,
\\
g_2 =  \frac{(k -s_1 2b)(2c -s_1k^2)}{4(k -s_1 b)} -\frac{kb}{2}, 
\quad
g_3 = \frac{k}{2} , 
\\
\varpi = \frac{k}{8(k -s_1b)}\Big( s_1 9b^2 + s_1 \frac{k^2}{2} - 5 bk -9 c \Big),
\\
E = \frac{kb^2}{4(k -s_1b)} \Big( c - \frac{4k b}{3} + s_1\frac{b^2}{3} + s_1\frac{k^2}{2} \Big),
\\
\begin{aligned}
\mu = & 8k(k - 2s_1 b)^2 \Big(\frac{2a}{k} +\frac{5(k b -2 c)}{4(k - s_1 b)}\Big)
\Big(\frac{2a}{k} +\frac{5(k b -2 c)}{8(k - s_1 b)}\Big)^2
\Big(\frac{2a}{k} +\frac{k b -2 c}{4(k - s_1 b)}\Big)^2 , 
\end{aligned}
\label{mu.pis2.fam1}
\end{gather}
\end{subequations}
which are subject to the conditions \eqref{Hsq.conds}--\eqref{V.conds}. 

The first two of the three conditions \eqref{V.conds} yield
\begin{equation}\label{signs1.pis2.fam1}
s_3=\sgn(k)
\end{equation}
and
\begin{equation}\label{ineqn1.pis2.fam1}
k(k - s_1 b)(2c -4 k b + s_1(k^2 + 2b^2)>0 , 
\end{equation}
while the third condition gives
\begin{equation}\label{ineqn2.C2sqnot0.pis2.fam1}
k(k - s_1 b)(6 k b - 18 c + s_1(k^2 + 2b^2)> 0 
\quad\text{when $\mu\neq0$},
\end{equation}
and
\begin{equation}\label{ineqn2.C2sqis0.pis2.fam1}
\begin{gathered}
(k - s_1 2 b) (2 b^2 + k^2 +s_1 (b k  - 8 c)) (2 b^2 + k^2 -s_1(2b k + 2c))  
(2 b^2 + k^2 +s_1 (6 b k -18 c))
=0 \\
\quad\text{when $\mu=0$}.
\end{gathered}
\end{equation}

In the case $\mu\neq0$, 
the first of the two conditions \eqref{Hsq.conds.C2sqnot0} 
combined with inequality \eqref{ineqn1.pis2.fam1} 
yields $s_1 k >0$,
which gives
\begin{equation}\label{signs2.pis2.fam1}
s_1=\sgn(k) . 
\end{equation}
The second of the two conditions \eqref{Hsq.conds.C2sqnot0} consists of
\begin{equation}\label{ineqn3.pis2.fam1}
(s_1 2 b -k)(k b - 8 c + s_1(k^2 + 2b^2))(2k b + 2 c - s_1(k^2 + 2b^2)) > 0 . 
\end{equation}
Inequalities \eqref{ineqn3.pis2.fam1}, \eqref{ineqn2.C2sqnot0.pis2.fam1}, \eqref{ineqn1.pis2.fam1}
are straightforward to solve after splitting into cases $k>0$ and $k<0$. 
This yields
\begin{subequations}\label{ineqns.pis2.fam1}
\begin{equation}\label{kpos.ineqn.pis2.fam1}
b < k/2,
\quad
-b^2 + 2k b - k^2/2 < c < b^2/9 + k b/3 + k^2/18, 
\quad
\text{ if } k>0 , 
\end{equation}
and
\begin{equation}\label{kneg.ineqn.pis2.fam1}
\begin{aligned}
b > |k|,
&\quad
-b^2/9 +k b/3 -k^2/18 < c<b^2+2 k b +k^2/2,
\\
|k|/2 < b < |k|,
&\quad
b^2 + 2 k b + k^2/2 < c < -b^2/9 + k b/3 - k^2/18,
\end{aligned}
\quad
\text{ if } k<0 . 
\end{equation}
\end{subequations}

For the case $\mu=0$, 
conditions \eqref{ineqn2.C2sqis0.pis2.fam1} and \eqref{ineqn1.pis2.fam1}
can be split into two subcases:
\begin{equation}\label{subcase1.C2sqis0.pis2.fam1}
b= s_1 k/2,
\quad
c> s_1 k^2/4
\end{equation}
and
\begin{equation}\label{c.subcase2.C2sqis0.pis2.fam1}
c= \nu_1 bk + s_1 \nu_2 (2b^2 +k^2),
\quad
(\nu_1,\nu_2) = (1/8,1/8), (1/3,1/18), (-1,1/2) , 
\end{equation}
\vskip -18pt
\begin{subequations}\label{b.subcase2.C2sqis0.pis2.fam1}
\begin{gather}
b < s_1 k/2 
\quad\text{ if }
k>0 , 
\\
b > s_1 k/2 
\quad\text{ if }
k<0 . 
\end{gather}
\end{subequations}

In the first subcase \eqref{subcase1.C2sqis0.pis2.fam1},
conditions \eqref{Hsq.conds.C2sqis0} reduce to $s_1 k<0$,
which yields
\begin{equation}\label{sign1.subcase1.C2sqis0.pis2.fam1}
s_1 = -\sgn(k) . 
\end{equation}
There are no additional conditions. 

In the second subcase \eqref{c.subcase2.C2sqis0.pis2.fam1}--\eqref{b.subcase2.C2sqis0.pis2.fam1}, 
conditions \eqref{Hsq.conds.C2sqis0} consist of
\begin{subequations}\label{ineqns.subcase2.C2sqis0.pis2.fam1}
\begin{align}
& k\leq s_1 2b 
\quad\text{ for } (\nu_1,\nu_2) = (1/8,1/8) , 
\\
& k\geq s_1 2b 
\quad\text{ for } (\nu_1,\nu_2) = (1/3,1/18) , 
\end{align}
while when $(\nu_1,\nu_2) = (-1,1/2)$ they hold as equalities. 
For the latter, the further condition \eqref{Hsq.muis0.conds} is necessary,
which yields $s_1 k>0$ and thus
\begin{equation}
s_1 =\sgn(k)
\quad\text{ for } (\nu_1,\nu_2) = (-1,1/2) . 
\end{equation}
\end{subequations}
These inequalities \eqref{ineqns.subcase2.C2sqis0.pis2.fam1} 
along with inequalities \eqref{b.subcase2.C2sqis0.pis2.fam1}
finally give 
\begin{subequations}\label{conds.subcase2.C2sqis0.pis2.fam1}
\begin{align}
& s_1=1,
\quad
b> -|k|/2, 
\quad
k<0, 
\quad\text{ when }
(\nu_1,\nu_2) = (1/8,1/8), 
\\
& s_1 =1,
\quad
b<k/2,
\quad
k>0,
\quad\text{ when }
(\nu_1,\nu_2) = (1/3,1/18), (-1,1/2) , 
\end{align}
\end{subequations}
and
\begin{subequations}
\begin{align}
& s_1 =-1,
\quad
b>|k|/2,
\quad
k<0,
\quad\text{ when }
(\nu_1,\nu_2) = (1/3,1/18), (-1,1/2) , 
\\
& s_1 =-1,
\quad
b<-k/2,
\quad
k>0,
\quad\text{ when }
(\nu_1,\nu_2) = (1/8,1/8) . 
\end{align}
\end{subequations}

The \underline{second family} has two free parameters $\varpi$, $k\neq0$, 
The determined parameters and signs are given by 
%maple pis2-case2
\begin{subequations}\label{params.pis2.fam2}
\begin{gather}
s_2=1, 
\quad
a =-\frac{13c + 4\varpi}{36},
\quad
h = -s_3 s_1 \frac{13c + 4\varpi}{3 b},
\\
g_1 = -s_1\Big( \frac{61 b^3}{81} + \frac{64 \varpi^2}{81b}\Big) -\frac{127 b\varpi}{81}, 
\quad
g_2 = s_1\frac{2 b^2}{9} +\frac{4\varpi}{9},
\quad
g_3 = s_1\frac{b}{2}, 
\\
\mu = s_1 \frac{2(2c + \varpi)}{c}\Big( \frac{(c +\varpi)(59c +32\varpi)}{81}\Big)^2,
\quad
E = -s_1\frac{c(7c +4\varpi)}{9},
\\
c = s_1 \frac{b^2}{2},
\quad
b = s_1 k , 
\label{kb.pis2.fam2}
\end{gather}
\end{subequations}
which are subject to the conditions \eqref{Hsq.conds}--\eqref{V.conds}. 

Relation \eqref{kb.pis2.fam2} implies $\sgn(k) = s_1\sgn(b)$, 
and then the first of the three conditions \eqref{V.conds} yields
\begin{equation}
s_3=\sgn(k)
\end{equation}
The remaining two conditions give 
\begin{equation}\label{Vcond1.pis2.fam2}
13 s_1 b^2 + 8 \varpi < 0
\end{equation}
in addition to 
\begin{equation}\label{Vcond2.C2sqnot0.pis2.fam2}
s_1 b^2 +\varpi > 0
\quad\text{when $\mu\neq0$}, 
\end{equation}
and 
\begin{equation}\label{Vcond2.C2sqis0.pis2.fam2}
(s_1 b^2 +\varpi)(s_1 b^2 +2\varpi)(59 s_1 b^2 +34\varpi)=0
\quad\text{when $\mu=0$}. 
\end{equation}

In the case $\mu\neq0$, 
the first of the two conditions \eqref{Hsq.conds.C2sqnot0} reduces to the inequality
\begin{equation}\label{b.C2sqnot0.pis2.fam2}
b>0, 
\end{equation}
whence 
\begin{equation}
s_1=\sgn(k)
\end{equation}
from relation \eqref{kb.pis2.fam2}. 
The second condition then reduces to 
\begin{equation}\label{Hsqcond.C2sqnot0.pis2.fam2}
s_1(s_1 b^2 +2\varpi)(s_1 59 b^2 + 64 \varpi) < 0
\end{equation}
It is straightforward to solve inequalities \eqref{Hsqcond.C2sqnot0.pis2.fam2}
and \eqref{Vcond1.pis2.fam2}--\eqref{Vcond2.C2sqnot0.pis2.fam2}
after splitting into cases $s_1=\pm 1$. 
There is no solution for $s_1=1$,
whereas $s_1=-1$ gives 
\begin{equation}\label{ineqns.pis2.fam2.C2sqnot0}
b^2 < \varpi < 13b^2/8
\end{equation}
This implies  
\begin{equation}\label{signs.pis2.fam2.C2sqnot0}
s_1 = -1, 
\quad
s_3 = -1,
\quad
k<0
\end{equation}

In the case $\mu=0$, 
conditions \eqref{Hsq.conds.C2sqis0} consist of 
\begin{equation}\label{ineqns.C2sqis0.pis2.fam2}
s_1 b (s_1 b^2+\varpi)(s_1 b^2+2\varpi)^2 \geq 0,
\quad
s_1 b (s_1 59b^2+64\varpi)(s_1 b^2+2\varpi)^2 \leq 0
\end{equation}
These inequalities together with the previous conditions \eqref{Vcond1.pis2.fam2} and \eqref{Vcond2.C2sqis0.pis2.fam2}
are straightforward to solve after splitting into cases $s_1=\pm 1$. 
This yields three subcases:
\begin{subequations}\label{b.subcases.pis2.fam2}
\begin{align}
& s_1 = -1,
\quad
b<0,
\quad
\varpi = 59 b^2/64
\label{b.subcase2.pis2.fam2}\\
& s_1 = -1,
\quad
b>0,
\quad
\varpi = b^2
\label{b.subcase3.pis2.fam2}
\end{align}
and also $s_1=-1$, $\varpi =b^2/2$, $b\gtrless 0$. 
In the latter subcase, conditions \eqref{ineqns.C2sqis0.pis2.fam2} hold as equalities, 
and this requires imposing the further condition \eqref{Hsq.muis0.conds}, 
which yields
\begin{equation}\label{b.subcase1.pis2.fam2}
s_1 = -1,
\quad
b>0,
\quad
\varpi =b^2/2
\end{equation}
\end{subequations}

Combining the relations \eqref{b.subcases.pis2.fam2} with expression \eqref{kb.pis2.fam2} then gives
\begin{subequations}\label{k.pis2.fam2}
\begin{align}
& k>0 
\quad\text{ for }
\varpi = 59 b^2/64
\\
& k<0 
\quad\text{ for }
\varpi = b^2, b^2/2
\end{align}
\end{subequations}

Note that the subcase $b^2=\varpi$ for $\mu=0$ can be merged with 
the subcase $b^2<\varpi$ for $\mu>0$: $b^2\leq \varpi$ for $\mu\geq 0$.

{\bf Solution family for $\boldsymbol{p=3}$}:

This family has two free parameters $b\neq0$, $k\neq0$.
The determined parameters and signs are given by 
%maple pis3
\begin{subequations}\label{params.pis3}
\begin{gather}
s_1=1,
\quad
a = -\frac{k(k+3b^2)}{40b},
\quad
h = -s_3\frac{k+3b^2}{2b},
\label{gh.pis3}
\\
g_1 = -s_2 \frac{75b^6 -220 k b^4 + 182 k^2 b^2 -30 k^3}{300 b^2}, 
\quad
g_2 =  -s_2 \frac{(2k-b^2)(3k-5b^2)}{45 b}, 
\quad
g_3 = s_2 \frac{3k}{10} , 
\label{gs.pis3}
\\
E=-\frac{k b(k+b^2)}{20}, 
\quad
\mu = \frac{4k(2k -b^2)(k-5b^2)^2(3k-5b^2)^2(15k- 11b^2)^2}{3(15b)^5}, 
\\
\varpi=\frac{2k(k -3b^2)}{5b},
\quad
c= -\frac{21k^2 -53kb^2 +10b^4}{90b}
\end{gather}
\end{subequations}
which are subject to the conditions \eqref{Hsq.conds}--\eqref{V.conds}. 

The first of the three conditions \eqref{V.conds} gives 
\begin{equation}\label{s3.pis3}
s_3=\sgn(k)
\end{equation}
while the other two conditions reduce to 
$k b (k+ 3b^2) < 0$ 
in addition to $k b (2k -b^2) >0$ for $\mu\neq0$,
and $(2k -b^2)(k-5b^2)(3k-5b^2)(15k- 11b^2)=0$ for $\mu=0$. 

In the case $\mu\neq0$, 
these inequalities yield 
\begin{equation}\label{kb.cond.muisnot0.pis3}
\begin{aligned}
& b<0, 
\quad
b^2> 2k
&&
\text{ if } 
k>0
\\
& b>0, 
\quad
b^2 >  -k/3
&&
\text{ if } 
k<0
\end{aligned}
\end{equation}
whereas in the case $\mu=0$ they give
\begin{equation}\label{kb.cond.muis0.pis3}
k>0, 
\quad
b<0, 
\quad
b^2 = 2k, k/5, 3k/5, 15k/11
\end{equation}

Conditions \eqref{Hsq.conds.C2sqnot0} give
\begin{equation}\label{s2.muisnot0.pis3}
s_2 = -1
\quad\text{when $\mu\neq0$}. 
\end{equation}
Conditions \eqref{Hsq.conds.C2sqis0} give
\begin{equation}\label{s2.muis0.subcase1&3.pis3}
s_2 = -1
\quad\text{ for } 
b^2 = 2k, 3k/5
\quad\text{when $\mu=0$} , 
\end{equation}
while for $b^2=k/5,15k/11$, 
conditions \eqref{Hsq.conds.C2sqis0} hold as equalities. 
Hence, the stronger conditions \eqref{Hsq.muis0.conds} need to be considered. 
Substitution of expressions \eqref{gh.pis3}, \eqref{gs.pis3}, and \eqref{s3.pis3}
into those conditions, followed by evaluation for the two values of $b$, 
yields $s_2 k>0$. 
Since $k>0$, 
this gives 
\begin{equation}\label{s2.muis0.subcase2&4.pis3}
s_2 = 1
\quad\text{ for } 
b^2 = k/5, 15k/11
\quad\text{when $\mu=0$}. 
\end{equation}

Note that the case $b^2=2k$ for $\mu=0$ can be merged with 
the case $b^2>2k$ for $\mu=0$: $b^2\geq 2k$ for $\mu\geq 0$.

{\bf Solution family for $\boldsymbol{p=4}$}:

This family has a single free parameter $k\neq0$.
The determined parameters and signs are given by 
%maple pis4
\begin{subequations}\label{params.pis4}
\begin{gather}
s_2=1,
\quad
a =-\frac{kb}{10},
\quad
h = -s_3\frac{3b}{2}, 
\\
g_1 = -\frac{18 k b^2}{25}, 
\quad
g_2 = \frac{3k b}{10}, 
\quad
g_3 = \frac{k}{5}, 
\quad
E = s_1\frac{k^2}{15},
\\
\mu = s_1 2\Big(\frac{81k^2 b}{100}\Big)^2,
\label{mu.pis4}
\\
\varpi = -\frac{5k b}{4},
\quad
c = \frac{k b}{2},
\\
2 k + s_1 b^3 =0, 
\label{kb.pis4}
\end{gather}
\end{subequations}
which are subject to the conditions \eqref{Hsq.conds}--\eqref{V.conds}. 

Relation \eqref{kb.pis4} implies $\sgn(k) = -s_1\sgn(b)$, with $b\neq 0$.
Then, conditions \eqref{V.conds} show that
\begin{equation}
s_1 =1,
\quad
s_3=\sgn(k) 
\end{equation}
where $\sgn(k)=-\sgn(b)\neq0$
and hence $\mu>0$ from expression \eqref{mu.pis4}. 
Substitution of these signs into conditions \eqref{Hsq.conds.C2sqnot0} 
yields $b^5<0$. 
This gives
\begin{equation}
b<0,
\quad 
k>0 
\end{equation}

\end{document}